\documentclass[a4paper,11pt]{article}
\pdfoutput=1
\usepackage{jheppub}
\usepackage{amsfonts,amssymb,amsmath}
\usepackage[normalem]{ulem}
\usepackage{color}
\usepackage{mathtools}
\usepackage{epstopdf}
\usepackage{tensor}
\usepackage{enumerate}
\usepackage{pdflscape}
\allowdisplaybreaks

\newcommand{\beqa}{\begin{eqnarray}}
\newcommand{\eeqa}{\end{eqnarray}}
\newcommand{\beqar}{\begin{eqnarray*}}
\newcommand{\eeqar}{\end{eqnarray*}}
\newcommand{\reef}[1]{(\ref{#1})}

\newcommand{\be}{\begin{equation}}
\newcommand{\ee}{\end{equation}}
\newcommand{\ba}{\begin{eqnarray}}
\newcommand{\ea}{\end{eqnarray}}
\newcommand{\beal}{\begin{aligned}}
\newcommand{\eeal}{\end{aligned}}
\newcommand{\nn}{\nonumber}

\newcommand\cO{{\cal O}}

\title{Quintessential Quartic Quasi-topological Quartet }

\author[a, c]{Jamil Ahmed}
\author[a]{Robie A. Hennigar}
\author[a,b]{Robert B. Mann}
\author[a, d]{Mozhgan Mir}

\affiliation[a]{Department of Physics and Astronomy, University of Waterloo,
Waterloo, Ontario, Canada, N2L 3G1}
\affiliation[b]{Perimeter Institute, 31 Caroline Street North, Waterloo,
ON, N2L 2Y5, Canada}
\affiliation[c]{Department of Mathematics, Quaid-i-Azam University, Islamabad, Pakistan}

\affiliation[d]{School of Physics, Institute for Research in Fundamental Sciences (IPM), P.O.Box 19395-5531, Tehran, Iran}

\emailAdd{jahmed@student.qau.edu.pk}
\emailAdd{rhennigar@uwaterloo.ca}
\emailAdd{rbmann@uwaterloo.ca}
\emailAdd{mmir@ipm.ir}

\abstract{
We construct the quartic version of generalized quasi-topological gravity, which was recently constructed to cubic order in arXiv: 1703.01631. This class of theories includes Lovelock gravity and a known form of quartic
quasi-topological gravity as special cases and possess a number of remarkable properties: (i) In vacuum, or in the presence of suitable matter, there is a single independent field equation which is a total derivative.  (ii) At the linearized level, the equations of motion on a maximally symmetric background are second order, coinciding with the linearized Einstein equations up to a redefinition of Newton's constant.  Therefore, these theories propagate only the massless, transverse graviton on a maximally symmetric background. (iii) While the Lovelock and quasi-topological terms are trivial in four dimensions, there exist four new generalized quasi-topological terms (the quartet) that are nontrivial, leading to interesting higher curvature theories in $d \geq 4$ dimensions that appear well suited for holographic study.  We construct four dimensional black hole solutions to the theory and study their properties.  A study of black brane solutions in arbitrary dimensions reveals that these solutions are modified from the `universal' properties these solutions have.  This result may lead to interesting consequences for the dual CFTs.
}

\preprint{DCPT-17/03}

\begin{document}

\maketitle

\section{Introduction}

It is generally expected that in a quantum theory of gravity the Einstein-Hilbert action will be modified by the addition of higher curvature terms.  Within the context of string theory, these appear in the $\alpha'$ corrections to the low energy effective action, including the Gauss-Bonnet term, which falls into the Lovelock class~\cite{Lovelock:1971yv}, and various higher order corrections which have been computed by various authors~\cite{Zwiebach:1985uq, Metsaev:1986yb, Gross:1986mw, Myers:1987qx}.

More pragmatically, higher curvature gravity is interesting in its own right.  It has been known for more than forty years that these theories allow for renormalizable quantum gravity~\cite{Stelle:1976gc}.
In the context of holography, the study of higher curvature toy models has led to the discovery of numerous interesting properties---some universal~\cite{Bueno:2015rda}---of conformal field theories~\cite{Camanho:2009vw, Buchel:2009sk, Camanho:2009hu,Hofman:2009ug, Brigante:2007nu}. For example, the inclusion of quadratic terms has been shown to lead to violations of  the Kovtun--Son--Starinets (KSS) viscosity/entropy ratio bound~\cite{Kovtun:2004de, Brigante:2007nu} and studies of cubic curvature theories have led to holographic $c$-theorems~\cite{Zamolodchikov:1986gt}, valid in arbitrary dimensions~\cite{Myers:2010tj}.  Thermodynamic considerations reveal that black holes in higher curvature theories have nontrivial behaviour, giving rise to isolated critical points and superfluid-like phase transitions~\cite{  Hennigar:2016xwd, Dolan:2014vba, Kubiznak:2016qmn}.

Holographic considerations partially motivated the construction of a new cubic theory of gravity, \textit{quasi-topological gravity}~\cite{Oliva:2010eb, Oliva:2010zd, Myers:2010ru, Myers:2010jv, Cisterna:2017umf, Ghodsi:2017iee, Dehghani:2011hm, Dehghani:2011vu, Dehghani:2013ldu}, which possesses a number of remarkable properties. While the cubic Lovelock term---the six dimensional Euler density---is gravitationally non-trivial only in $d > 6$, the cubic quasi-topological term contributes to the field equations in five dimensions and higher.  The equations of motion, which are fourth order on general backgrounds, reduce to second order  under the restriction to spherical symmetry.  The theory admits exact spherically symmetric black hole solutions with the metric function determined by a polynomial equation very similar to the Wheeler polynomial of Lovelock gravity.  Remarkably, despite the field equations being fourth order on general backgrounds, the linearized equations of motion describing graviton propagation in a maximally symmetric background are second order and match the linearized Einstein equations, up to a redefinition of Newton's constant~\cite{Myers:2010ru, Bueno:2016ypa}.  In other words, the additional massive scalar mode and massive, ghost-like graviton are absent. The upshot of this is that quasi-topological gravity avoids an unpalatable feature that afflicts many higher curvature theories: the propagation of ghosts and tachyons in the vacuum.   

In a recent paper~\cite{Hennigar:2017ego} two of us have shown that cubic quasi-topological gravity and cubic Lovelock gravity can be understood as members of a class of gravitational theories---\textit{generalized quasi-topological gravity}---which, under the restriction of spherical symmetry, have a single independent field equation.  This is a sufficient condition to allow vacuum static spherically symmetric (VSSS) solutions described by a single metric function; that is, solutions of the form
\be\label{eqn:metricAnsatz}
ds^2 = - N^2 f dt^2 + \frac{dr^2}{f} + r^2 d \Sigma^2_{(d-2), k}\,,
\ee
with $N = const.$, i.e., the solution  is characterized in terms of a single metric function $f$~\cite{Jacobson:2007tj}.\footnote{Henceforth  we normalize $N$ to unity, setting $N=1$. This can be achieved without loss of generality by reparametrizing the time coordinate $t$.} Here $d \Sigma^2_{(d-2), k}$ is the line element on a surface of constant scalar curvature $k = +1,0, -1$ corresponding to spherical, flat, and hyperbolic topologies.  In~\cite{Hennigar:2017ego} we demonstrated that the most general theory to cubic order in curvature  having this property  is given by the action
\be\label{eqn:actionWithS}
\mathcal{I} = \frac{1}{16 \pi G} \int d^d x \sqrt{-g} \bigl[-2 \Lambda + R + \alpha \mathcal{X}_4+  \beta \mathcal{X}_6 + \mu \mathcal{Z}_d -  \lambda \mathcal{S}_d \bigr]\,.
\ee
Here, $\Lambda$ is the cosmological constant and $\alpha,\beta,\mu, \lambda$ are arbitrary coupling constants. $R$ stands for the Ricci scalar and $\mathcal{X}_4$ and $\mathcal{X}_6$ are the four- and six-dimensional Euler densities,
\ba\label{X6}
\mathcal{X}_4&=&-\frac{1}{4}\delta^{a_1b_1a_2b_2}_{c_1d_1c_2d_2}R_{a_1b_1}{}^{c_1d_1}R_{a_2b_2}{}^{c_2d_2}\,,\nonumber\\
\mathcal{X}_6  &=& - \frac{1}{8}
\delta^{a_1b_1a_2b_2a_3b_3}_{c_1d_1c_2d_2c_3d_3}R_{a_1b_1}{}^{c_1d_1}R_{a_2b_2}{}^{c_2d_2}R_{a_3b_3}{}^{c_3d_3}\,,
\ea
corresponding to the standard Gauss-Bonnet and cubic Lovelock terms, respectively. $\mathcal{Z}_d$ is the cubic quasi-topological term given by \eqref{Zd} below, and $\mathcal{S}_d$ is a new term whose  explicit form 
\ba\label{Sd}
\mathcal{S}_d &=&
14 R_{a}{}^{e}{}_{c}{}^{f} R^{abcd} R_{bedf}+ 2 R^{ab} R_{a}{}^{cde} R_{bcde}- \frac{4 (66 - 35 d + 2 d^2) }{3 (d-2) (2 d-1)} R_{a}{}^{c} R^{ab} R_{bc}\nonumber\\
&& -  \frac{2 (-30 + 9 d + 4 d^2) }{(d-2) (2 d-1)} R^{ab} R^{cd} R_{acbd} -  \frac{(38 - 29 d + 4 d^2)}{4 (d -2) (2 d  - 1)} R R_{abcd} R^{abcd}  \nonumber\\
&&+ \frac{(34 - 21 d + 4 d^2) }{(d-2) ( 2 d - 1)} R_{ab} R^{ab} R -  \frac{(30 - 13 d + 4 d^2)}{12 (d-2) (2 d - 1)}  R^3 
\ea
was elucidated for the first time in~\cite{Hennigar:2017ego}.  Interestingly, while both the cubic Lovelock and quasi-topological terms vanish in four dimensions, the new term $\mathcal{S}_4$ makes a non-trivial contribution to the field equations, reducing to the contribution from \textit{Einsteinian cubic gravity}~\cite{Bueno:2016xff}.  However, while Einsteinian cubic gravity does not permit solutions of the form~\eqref{eqn:metricAnsatz} in $d > 4$, $\mathcal{S}_d$ does.  In this sense, $\mathcal{S}_d$ can be viewed as the $d$-dimensional generalization of the four-dimensional Einsteinian cubic term.  

In~\cite{Hennigar:2017ego} it was observed that the linearized equations of motion derived from the action~\eqref{eqn:actionWithS} coincide with the linearized Einstein equations, up to a redefinition of Newton's constant.  Thus, to cubic order in curvature, the entire class of theories which have a single independent field equation for a VSSS ansatz enjoy the property of propagating only the massless, transverse graviton familiar from Einstein gravity.  In~\cite{Hennigar:2017ego} it was conjectured that this would be a general feature for this class of theories to all orders in the curvature.  Shortly after this, it was demonstrated in~\cite{Bueno:2017sui} that this is indeed the case for any theory for which the metric~\eqref{eqn:metricAnsatz} describes the gravitational field outside a spherically symmetric mass distribution.  This caveat explains why some theories, such as $f(R)$ gravity,    admit solutions of the form~\eqref{eqn:metricAnsatz} with $N = 1$ but also propagate additional modes on the vacuum: in these theories, the metric~\eqref{eqn:metricAnsatz} does not describe the gravitational field of a spherical mass~\cite{Bueno:2017sui}.

The aim of the present paper is to provide the quartic version of generalized quasi-topological gravity: describing all quartic Lagrangian densities which, under the restriction to a VSSS ansatz, have a only a single independent field equation.  
We find a rather broad class of interesting theories, including five new quasi-topological theories and several quartic generalizations
of the cubic Lagrangian \eqref{Sd}.

Our paper is organized as follows.  In section~\ref{sec:construction} we first review the procedure by which the generalized quasi-topological gravities can be constructed, and present the results of this construction for the quartic case.  In section~\ref{sec:linear} we discuss the linearized theory and in section \ref{sec:fieldeqns} we derive the field equations   from the actions we construct. Finally, in section~\ref{sec:blackholes} we present the contributions to the Wald entropy arising from each of the interactions presented, and we present four-dimensional black hole solutions of this theory.  We go on to study black brane solutions in arbitrary dimensions, finding they satisfy the expected relations of a CFT living in one dimension fewer.

\section{Construction of the quartic theories}
\label{sec:construction}

\subsection{Review of the construction}

We begin by briefly reviewing the construction used to obtain the generalized quasi-topological theories in~\cite{Hennigar:2017ego}.  We refer the reader also to ref.~\cite{Bueno:2017sui} where this general strategy has recently been nicely discussed.  The central idea is to construct theories that supplement Einstein gravity with higher curvature terms in a manner such that these terms can be ``turned off" by a suitable adjustment of parameters in the action.  Our conditions are the same as those mentioned in~\cite{Bueno:2016lrh} and are, effectively, designed so that the most general solution of the theory takes the form of~\eqref{eqn:metricAnsatz} with $N = 1$.  Explicitly, the conditions are:
\begin{enumerate}
\item The solution is not an `embedding' of an Einstein gravity black hole into a higher order gravity~\cite{delaCruzDombriz:2009et, Smolic:2013gz, Lu:2012xu}.  That is, the solution must be modified by the addition of the higher curvature terms.

\item The solution is not of a pure higher order gravity, but includes the Einstein-Hilbert term.  For example, pure Weyl-squared gravity allows for four dimensional solutions with $N=1$~\cite{delaCruzDombriz:2009et, Riegert:1984zz, Oliva:2011xu, Oliva:2012zs}.

\item Further, the theory must admit an Einstein-Gravity limit, i.e. reduce to the Einstein-Hilbert action upon setting
some of the parameters in the action to zero.  This excludes certain theories that tune the couplings between the various orders of curvature terms \cite{Bueno:2016dol, Cai:2009ac}.
\end{enumerate}

A sufficient condition for this is that, upon setting $N = 1$ in the vacuum field equations, we are left with only a single independent field equation for any metric function $f(r)$.  That is, we demand that,
\be\label{eqn:feq_condition} 
\left(\mathcal{E}_t^t - \mathcal{E}_r^r \right) \big|_{N = 1} = 0 \, ,
\ee
where
\be 
\mathcal{E}_{ab} = \frac{1}{\sqrt{-g}} \frac{\delta \mathcal{I}}{\delta g^{ab}}
\ee 
is the generalized Einstein tensor.\footnote{Note that in the case of spherical symmetry, the angular components of the field equations are satisfied provided the time and radial components of the field equations are satisfied.  This is a consequence of the contracted generalized Bianchi identity, $\nabla^a \mathcal{E}_{ab} = 0$, which follows from the diffeomorphism invariance of the theory.} We emphasize once again that in enforcing Eq.~\eqref{eqn:feq_condition} we do not place any constraints on the metric function $f$.  In a general quartic theory, evaluating the field equations in full generality is an arduous task.  It is more convenient to enforce~\eqref{eqn:feq_condition} by taking advantage of the Weyl method~\cite{palais1979, Deser:2003up}.  Here, one inserts the metric ansatz~\eqref{eqn:metricAnsatz} into the action, integrates by parts to remove boundary terms, and varies the action with respect to $N$ and $f$ to obtain the two field equations.  A simple application of the chain rule reveals that
\be 
\frac{\delta \mathcal{I}}{\delta N} = \omega^{(k)}_{(d-2)} r^{d-2} \frac{2 \mathcal{E}_{tt}}{f N^2} \, , \quad  \frac{\delta \mathcal{I}}{\delta f} = -  \frac{\omega^{(k)}_{(d-2)} r^{d-2}}{f}\left[\mathcal{E}_t^t - N \mathcal{E}_r^r \right]
\ee 
and so   condition~\eqref{eqn:feq_condition} becomes
\be\label{eqn:action_condition} 
 \frac{\delta \mathcal{I}}{\delta f} \bigg|_{N = 1} = 0
\ee 
as was pointed out in~\cite{Bueno:2017sui}.  

Carrying out this procedure for a cubic theory of gravity, one is led to the following action
\be
\mathcal{I} = \frac{1}{16 \pi G} \int d^d x \sqrt{-g} \bigl[-2 \Lambda + R + \alpha \mathcal{X}_4+  \beta \mathcal{X}_6 + \mu \mathcal{Z}_d -  \lambda \mathcal{S}_d \bigr]\,,
\ee
where $\Lambda$ is the cosmological constant and $\alpha,\beta,\mu, \lambda$ are arbitrary coupling constants.  $R$ stands for the Ricci scalar and  $\mathcal{X}_4$ and $\mathcal{X}_6$ are the four- and six-dimensional Euler densities, corresponding to the standard Gauss-Bonnet and cubic Lovelock terms, respectively.  
$\mathcal{Z}_d$ is the cubic quasi-topological term,
\ba
\mathcal{Z}_d &=&    {{{R_a}^b}_c{}^d} {{{R_b}^e}_d}^f {{{R_e}^a}_f}^c
               + \frac{1}{(2d - 3)(d - 4)} \Bigl( \frac{3(3d - 8)}{8} R_{a b c d} R^{a b c d} R  - \frac{3(3d-4)}{2} {R_a}^c {R_c}^a R \nonumber \\
              &&  - 3(d-2) R_{a c b d} {R^{a c b}}_e R^{d e} + 3d  R_{a c b d} R^{a b} R^{c d}
                + 6(d-2) {R_a}^c {R_c}^b {R_b}^a  + \frac{3d}{8} R^3 \Bigr)\,.\quad
\label{Zd}
\ea
and $\mathcal{S}_d$ is a new term,   written explicitly in \eqref{Sd}.

Here we are interested in constructing the quartic generalization of this action.  We consider the following basis of quartic invariants~\cite{Bueno:2016ypa, Fulling:1992vm}:
\begin{align}
  \mathcal{L}_{1} &= R_{a}{}^{e}{}_{c}{}^{f} R^{abcd} R_{e}{}^{j}{}_{b}{}^{h} R_{fjdh} 
\,, \quad 
\mathcal{L}_{2}=R_{a}{}^{e}{}_{c}{}^{f} R^{abcd} R_{bjdh} R_{e}{}^{j}{}_{f}{}^{h}
\,, \quad
\mathcal{L}_{3}= R_{ab}{}^{ef} R^{abcd} R_{c}{}^{j}{}_{e}{}^{h} R_{djfh}
\,, \quad
\nonumber\\ 
\mathcal{L}_{4}&=R_{ab}{}^{ef} R^{abcd} R_{ce}{}^{jh} R_{dfjh}
\,,\quad
\mathcal{L}_{5}=R_{ab}{}^{ef} R^{abcd} R_{cdjh} R_{ef}{}^{jh}
\,, \quad
\mathcal{L}_{6}= R_{abc}{}^{e} R^{abcd} R_{fhjd} R^{fhj}{}_{e}
\,, 
\nonumber\\
\mathcal{L}_{7}&=(R_{abcd} R^{abcd})^2
\,,\quad
\mathcal{L}_{8}= R^{ab} R_{c}{}^{h}{}_{ea} R^{cdef} R_{dhfb}
\,, \quad
\mathcal{L}_{9}=R^{ab} R_{cd}{}^{h}{}_{a} R^{cdef} R_{efhb} 
\,,
\nonumber\\
\mathcal{L}_{10}&=R^{ab} R_{a}{}^{c}{}_{b}{}^{d} R_{efhc} R^{efh}{}_{d}
\,, \quad
\mathcal{L}_{11}= R R_{a}{}^{c}{}_{b}{}^{d} R_{c}{}^{e}{}_{d}{}^{f} R_{e}{}^{a}{}_{f}{}^{b}
\,, \quad 
\mathcal{L}_{12}=R R_{ab}{}^{cd} R_{cd}{}^{ef} R_{ef}{}^{ab},
\,
\nonumber\\
\mathcal{L}_{13}&= R^{ab} R^{cd} R_{ebfd} R^{e}{}_{a}{}^{f}{}_{c}
\,, \quad
\mathcal{L}_{14}=R^{ab} R^{cd} R_{ecfd} R^{e}{}_{a}{}^{f}{}_{b}
\,, \quad
\mathcal{L}_{15}= R^{ab} R^{cd} R_{efbd} R^{ef}{}_{ac}
\,,
\nonumber\\
\mathcal{L}_{16}&= R^{ab} R_{b}{}^{c} R_{defc} R^{def}{}_{a}
\,, \quad 
\mathcal{L}_{17}=R_{ef} R^{ef} R_{abcd} R^{abcd}
\,,\quad
\mathcal{L}_{18}=R^{de} R R_{abcd} R^{abc}{}_{e}
\,, 
\nonumber\\
\mathcal{L}_{19}&= R^2 R_{abcd} R^{abcd}
\, , \quad  
\mathcal{L}_{20}=R^{ab} R_{e}{}^{d} R^{ec} R_{acbd}
\,, \quad
\mathcal{L}_{21}=R^{ac} R^{bd} R R_{abcd}
\, ,  
\nonumber\\ 
\mathcal{L}_{22}&=R_{a}{}^{b} R_{b}{}^{c} R_{c}{}^{d} R_{d}{}^{a}
\,, \quad
\mathcal{L}_{23}=(R_{ab} R^{ab})^2 \,, \quad 
\mathcal{L}_{24}=R_{a}{}^{b} R_{b}{}^{c} R_{c}{}^{a} R
\,,
\nonumber\\
\mathcal{L}_{25}&=R_{ab} R^{ab} R^2
\,, \quad  \mathcal{L}_{26}=R^4 \, .
\end{align}
It is worth noting that in dimensions less than eight, the above 26 curvature invariants are not independent.  The reason is because a certain linear combination of these yields the eight dimensional Euler density,
\be 
\mathcal{X}_8 = \frac{1}{2^4}  \delta^{a_1b_1a_2b_2a_3b_3a_4b_4}_{c_1d_1c_2d_2c_3d_3c_4d_4}R_{a_1b_1}{}^{c_1d_1}R_{a_2b_2}{}^{c_2d_2}R_{a_3b_3}{}^{c_3d_3}R_{a_4b_4}{}^{c_4d_4}
\ee
which vanishes identically in dimensions less than eight.   Furthermore,  under the restriction to spherical symmetry, there are additional, subtle degeneracies.  There exist certain combinations of the above curvature invariants that  identically vanish for spherically symmetric metrics~\cite{Deser:2005pc}.  Thus, we can expect certain degeneracies of theories in the spherically symmetric case: The field equations will not change upon the addition of one of these terms to the action.  However, we should note that the resulting theories \textit{will be different} when one moves away from spherical symmetry.

In what follows we focus on the quartic contributions to the action and write the following action
\be\label{eqn:quartic_action_general} 
\mathcal{I} = \frac{1}{16 \pi G} \int d^d x \sqrt{-g} \left[-2 \Lambda + R + \sum_{i=1}^{26} c_i \mathcal{L}_i \right] 
\ee
turning off the quadratic and cubic terms for the time being.
In the following subsections we will enforce condition~\eqref{eqn:action_condition} on this theory by fixing the constants $c_i$ such that the condition is satisfied for any metric function $f$.  From a practical perspective, we first compute the action~\eqref{eqn:quartic_action_general} in complete generality by explicitly determining each of the 26 terms in arbitrary dimensions for a VSSS ansatz.  This procedure is made significantly more manageable via a simple script used to determine the dimension dependence.  Our results have been cross-checked up to (in some cases) 19 dimensions. All subsequent calculations were then performed working directly with this completely general action.

We will split our discussion into two main parts, focusing first on the case of dimensions larger than four and then the four dimensional case separately.  As was the case in the cubic theory~\cite{Hennigar:2017ego}, the four dimensional case is somewhat special, while all other dimensions can be treated on equal footing.

\subsection{The case for dimensions larger than four} 
In five and higher dimensions, there are nine constraints that determine the class of theories with this property.  We, somewhat arbitrarily, solve the constraints for $c_{12}$, $c_{17}$, $c_{19}$, $c_{20}$, $c_{21}$, $c_{22}$, $c_{23}$, $c_{24}$ and $c_{25}$, yielding lengthy expressions that we have included in appendix~\ref{app:constraints}.  There are then 17 free parameters that we can adjust to find interesting quartic curvature terms. For organizational purposes, to classify these theories we will split them   into two convenient 
categories: theories in which the resulting field equation is the total derivative of a polynomial of the metric function, and theories
in which the field equation contains more than one derivative of the metric function.
 In each case we will make remarks about the field equations, but postpone a full discussion of the resulting field equations until section~\ref{sec:fieldeqns}.  Here our aim is to present a convenient basis for the 17 quartic theories.

\subsubsection{Lovelock and quasi-topological theories}

We begin by determining what additional constraints are required so that all terms in the action leading to more than one derivative in the field equations for the VSSS metric function $f$ (see section~\ref{sec:fieldeqns}) are eliminated.   The resulting field equation is then a total derivative of a polynomial in $f$.   We find that two additional constraints is the \textit{minimum number required} to eliminate these higher derivative terms in the action.  We have checked that the conclusions that follow do not intimately depend on which two constants are solved for in these two constraints.  
Choosing somewhat arbitrarily $c_9$ and $c_{15}$, we find
 
\begin{align}
c_9 &= -{\frac { 2\left( -8+2{d}^{4}-23d+39{d}^{2}-16{d}^{3}
 \right) }{ \left( 3d-4 \right)  \left( d-4 \right)  \left( 
{d}^{2}-6d+11 \right) }} c_1  
-{\frac { 2\left( 122+130{d}^{2}-207
d-37{d}^{3}+4{d}^{4} \right) }{ \left( 3d-4 \right) 
 \left( d-4 \right)  \left( {d}^{2}-6d+11 \right) }}c_2
\nn\\
&-{\frac { \left( 5d-7 \right)  \left( d-4 \right) }{ \left( 3d-4
 \right)  \left( {d}^{2}-6d+11 \right) }} c_3
-{\frac { 8\left( 82+82
{d}^{2}-21{d}^{3}+2{d}^{4}-139d \right)}{ \left( 3d-4
 \right)  \left( d-4 \right)  \left( {d}^{2}-6d+11 \right) }}c_4
\nn\\
& 
-{
\frac { 16\left( 82+82{d}^{2}-21{d}^{3}+2{d}^{4}-139d \right) }{ \left( 3d-4 \right)  \left( d-4 \right)  \left( {d}^{2}-6
d+11 \right) }} c_5
-{\frac { 4 \left( d-2 \right)  \left( d-3 \right) 
 \left( 4{d}^{2}-17d+16 \right) }{ \left( 3d-4 \right) 
 \left( d-4 \right)  \left( {d}^{2}-6d+11 \right) }} c_6
\nn\\
&
-{\frac { 32
 \left( d-1 \right)  \left( d-3 \right)  \left( d-2 \right) ^{2}}{ \left( 3d-4 \right)  \left( d-4 \right)  \left( {d}^{2}-6d+
11 \right) }} c_7
-{\frac { \left( d-4 \right) }{2 \left({d}^{2}-6 d+11 \right)}} c_8
 - {\frac { \left( d-3
 \right) d}{2 \left({d}^{2}-6 d+11 \right)}} c_{10} \, ,
\nn\\
c_{15} &= {\frac {8{d}^{7}-111{d}^{6}+570{d}^{5}-1190{d}^{4}+210{d}^{3
}+2725{d}^{2}-3308d+1024}{ \left( 3d-4 \right)  \left( d-4
 \right)  \left( {d}^{2}-6d+11 \right)  \left( {d}^{3}-7{d}^{2}+14
d-4 \right) }} c_1
\nn\\
& + {\frac {2 \left( d-1 \right)  \left( 8{d}^{6}-116{d}^{5}+689{d
}^{4}-2141{d}^{3}+3661{d}^{2}-3197d+988 \right) }{ \left( 3d-4
 \right)  \left( d-4 \right)  \left( {d}^{2}-6d+11 \right)  \left( {
d}^{3}-7{d}^{2}+14d-4 \right) }} c_2
\nn\\
&+ {\frac {13{d}^{5}-167{d}^{4}+781{d}^{3}-1615{d}^{2}+1396d-
384}{ \left( 3d-4 \right)  \left( {d}^{2}-6d+11 \right)  \left( {d
}^{3}-7{d}^{2}+14d-4 \right) }} c_3
\nn\\
&+ {\frac {8 \left(4{d}^{7}-70{d}^{6}+513{d}^{5}-2022{d}^{4}+4566{d}
^{3}-5760{d}^{2}+3557d-716 \right)}{ \left( 3d-4 \right)  \left( d-4
 \right)  \left( {d}^{2}-6d+11 \right)  \left( {d}^{3}-7{d}^{2}+14
d-4 \right) }} (c_4 + 2 c_5)
\nn\\
& + {\frac { 4\left( d-1 \right)  \left( 8{d}^{6}-116{d}^{5}+673{d
}^{4}-1966{d}^{3}+2983{d}^{2}-2148d+512 \right) }{ \left( 3d-4
 \right)  \left( d-4 \right)  \left( {d}^{2}-6d+11 \right)  \left( {
d}^{3}-7{d}^{2}+14d-4 \right) }} c_6
\nn\\
&+{\frac {32 \left( d-1 \right)  \left( 2{d}^{4}-15{d}^{3}+32{d}
^{2}-9d-4 \right)  \left( d-2 \right)  \left( d-3 \right) }{ \left( 
3d-4 \right)  \left( d-4 \right)  \left( {d}^{2}-6d+11 \right) 
 \left( {d}^{3}-7{d}^{2}+14d-4 \right) }} c_7
\nn\\
& + {\frac { \left( d-3 \right) ^{2} \left( {d}^{2}-6d+2 \right) }{
 \left( {d}^{2}-6d+11 \right)  \left( {d}^{3}-7{d}^{2}+14d-4
 \right) }} c_8 +{\frac { \left( d-4 \right)  \left( 3{d}^{3}-21{d}^{2}+37d-11
 \right) }{ \left( {d}^{2}-6d+11 \right)  \left( {d}^{3}-7{d}^{2}+
14d-4 \right) }} c_{10}
\nn\\
&   - {\frac {{d}^{3}-8{d}^{2}+19d-8}{ 2 \left({d}^{3}-7{d}^{2}+14d-4 \right)}} c_{13} -{\frac {d \left( d-3 \right) }{{d}^{3}-7{d}^{2}+14d-4}} c_{14} + {\frac { \left( d-1 \right)  \left( d-4 \right) }{{d}^{3}-7{d}^{2}+14d-4}} c_{16}
\label{eqn:QT_constraints1}
\end{align}

We can place two additional constraints to remove the factors containing more than two derivatives of $N(r)$ in the action.  Of course, these terms would vanish anyway  since the theory is constructed in such a way that $N = 1$ solves one of the field equations, but imposing these additional constraints renders the variational principle much less cumbersome.   The fact that it is possible in this class of theories to kill off the higher powers of derivatives of $N(r)$, combined with the fact that the field equations for the resulting theory are algebraic, is consistent with Conjecture 2 of~\cite{Bueno:2017sui}.
 Choosing  $c_{18}$ and $c_{26}$  for this task, we find
 \begin{align}
c_{18} &= \frac{1}{\left( 3d-4 \right)  \left( d
-4 \right)  \left( {d}^{2}-6d+11 \right)  \left( {d}^{3}-7{d}^{2}+
14d-4 \right)  \left( {d}^{3}-9{d}^{2}+26d-22 \right) } \times
\nn\\
&\times \big[-2( 2{d}^{10}-112640{d}^{2}+6558{d}^{6}+71315{
d}^{3}-2329{d}^{7}-16827{d}^{4}+447{d}^{8}
\nn\\
&-46{d}^{9}-6654{d}
^{5}+87822d-28032) c_1 -  4
 ( 2{d}^{10}+156501{d}^{2}+16490{d}^{6}
\nn\\
&-158736{d}^{3}-
3749{d}^{7}+107067{d}^{4}+562{d}^{8}-50{d}^{9}-50145{d}^{5}-
92828d+25270) c_2 
\nn\\
& - 2(d-4) ( 4{d}^{8}-92{d}^{7}+900{d}^{6}-4901{d}
^{5}+16264{d}^{4}-33711{d}^{3}+42690{d}^{2}
\nn\\
&-30290d+9280) c_3  - 16 ( {d}^{10}+81391{d}^{2}+10461
{d}^{6}-93331{d}^{3}-2292{d}^{7}
\nn\\
&+67198{d}^{4}+325{d}^{8}-27
{d}^{9}-32176{d}^{5}-39268d+7550 ) c_4  - 32 ( {d}^{10}+81391{d}^{2}
\nn\\
&+10461{d}^{6}
-93331{d}^{3}-2292{d}^{7}+67198{d}^{4}+325{d}^{8}-27{d}^{9}-
32176{d}^{5}
\nn\\
&-39268d+7550 ) c_5 - 8( 2{d}^{10}+88410{d}^{2}+15059{d}^{6}-106219
{d}^{3}-3597{d}^{7}
\nn\\
&+81940{d}^{4}+555{d}^{8}-50{d}^{9}-42522
{d}^{5}-42618d+9088 ) c_6 
\nn\\
&- 32 \left( d-2 \right)\left( d-3 \right)  ( {d}^{8}
-18{d}^{7}+137{d}^{6}-573{d}^{5}+1436{d}^{4}-2191{d}^{3}+
1884{d}^{2}
\nn\\
&-584d-164 ) c_7  + \tfrac{1}{2}(3d-4)(d-4)\left( d-3 \right)  ( 5{d}^{5}-60{
d}^{4}+281{d}^{3}-626{d}^{2}
\nn\\
&+632d-184 ) c_8  \big]
-{\frac 
{ \left( d-5 \right)  \left( {d}^{3}-6{d}^{2}+10d-6
 \right) }{ \left( {d}^{2}-6d+11 \right)  \left( {d}^{3}-7{d}^{2}+
14d-4 \right) }} c_{10}
\nn\\
&+{\frac {3 \left( d-1 \right)  \left( d
-3 \right) }{{d}^{3}-9{d}^{2}+26d-22}} c_{11}
-{\frac { \left( 
d-3 \right) ^{2}}{{d}^{3}-7{d}^{2}+14d-4}} c_{13}
+2{\frac {
\left( d-3 \right) }{{d}^{3}-7{d}^{2}+14d-4}}  c_{14}
\nn\\
&
-{\frac {2 \left( d-3 \right) ^{2}}{{d}^{3}-7{d}^{2}+14d-4}} c_{16}
\nn\\
c_{26} &= \frac{1}{\left( 3d-4 \right)  \left( d-2 \right) ^{3} \left( {d}^{2}-6d+11
 \right)  \left( {d}^{3}-9{d}^{2}+26d-22 \right)  \left( {d}^{2}-7
d+14 \right)  \left( d-4 \right)} \times
\nn\\
&\times \bigg[\frac{c_1}{24} ( 147{d}^{10}+340880{d}^{2}+2457{d}^{6}-20
{d}^{11}-475564{d}^{3}+6300{d}^{7}+317426{d}^{4}
\nn\\
&-927{d}^{8}-
366{d}^{9}+{d}^{12}-97214{d}^{5}-87616d-2048  ) + \frac{c_2}{12} (200{d}^{10
}-467516{d}^{2}
\nn\\
&-29192{d}^{6}-22{d}^{11}+359578{d}^{3}+3084{d
}^{7}-208682{d}^{4}+1757{d}^{8}-919{d}^{9}+{d}^{12}
\nn\\
&+94783{d}^{
5}+393168d-155456) + \frac{c_3}{12} (d-4)( {d}^{10}-20{d}^{9}+150
{d}^{8}-440{d}^{7}
\nn\\
&-401{d}^{6}+6292{d}^{5}-16150{d}^{4}+12280
{d}^{3}+13312{d}^{2}-26080d+10240)
\nn\\
&+ \frac{c_4}{6} (d-2)({d}^{11}-22{d}^{10}+194
{d}^{9}-806{d}^{8}+995{d}^{7}+4130{d}^{6}-13426{d}^{5}-20342
{d}^{4}
\nn\\
&+181192{d}^{3}-412060{d}^{2}+442000d-194240) + \frac{c_5}{3} (d-2)({d}^{11}-22{d}^{10}+194{d}^{9}
\nn\\
&-806{d}^{8}+995{d}^{7}+4130{d}
^{6}-13426{d}^{5}-20342{d}^{4}+181192{d}^{3}-412060{d}^{2}
\nn\\
&+  442000d-194240 ) + \frac{c_6}{6} (195{d}^{10}+499536{d}^{2}-13335{d}^{6}-22{d}^{
11}-403948{d}^{3}
\nn\\
&+3582{d}^{7}+163666{d}^{4}+1153{d}^{8}-826{
d}^{9}+{d}^{12}-14578{d}^{5}-316736d+79872 ) 
\nn\\
&+ \frac{c_7}{3} (d-2)({
d}^{11}-18{d}^{10}+122{d}^{9}-324{d}^{8}-169{d}^{7}+2302{d}^
{6}+810{d}^{5}-28868{d}^{4}
\nn\\
&+88832{d}^{3}-152480{d}^{2}+168128
d-87552 ) \bigg]
\nn\\
&-
{\frac { \left( 3{d}^{4}-28{d}^{3}+105{d}^{2}-176d+120
 \right)  \left( d-4 \right) ^{2}}{ 4 \left( {d}^{2}-7d+14
 \right)  \left( {d}^{3}-9{d}^{2}+26d-22 \right)  \left( {d}^{2}-6
d+11 \right)  \left( d-2 \right) ^{3}}} c_8
\nn\\
& -{\frac { \left( 
{d}^{4}-6{d}^{3}+8{d}^{2}+18d-24 \right) }{ 2\left( {d}^{2}-7d+
14 \right)  \left( {d}^{2}-6d+11 \right)  \left( d-2 \right) ^{3}}} c_{10}
\nn\\
&+
{\frac { \left( {d}^{5}-10{d}^{4}+29{d}^{3}+16{d
}^{2}-172d+152 \right) }{2 \left( {d}^{2}-7d+14 \right)  \left( {d}
^{3}-9{d}^{2}+26d-22 \right)  \left( d-2 \right) ^{3}}} c_{11}  +{
\frac { \left( d-4 \right) }{2 \left( {d}^{2}-7d+14 \right) 
 \left( d-2 \right) ^{3}}} c_{13}
\nn\\
&-{\frac { \left( {d}^{2}-6d+12
 \right) }{2 \left( {d}^{2}-7d+14 \right)  \left( d-2
 \right) ^{3}}} c_{14}
+{\frac { \left( d-4 \right) }{ \left( {d}^{2}
-7d+14 \right)  \left( d-2 \right) ^{3}}}c_{16}
\nn\\
&-{\frac { \left( {d}^{
3}-8{d}^{2}+20d-8 \right) }{2 \left( {d}^{2}-7d+14
 \right)  \left( d-2 \right) ^{3}}} c_{18}
 \label{eqn:QT_constraints2}
\end{align}

Such theories yield a field equation of the form appearing in   quartic Lovelock gravity or the  more general quartic quasi-topological gravity \cite{Dehghani:2011vu}: a total derivative of a polynomial in $f(r)$.
 However, although there are still 13 free parameters after the two additional constraints \eqref{eqn:QT_constraints1} are imposed,    given the constraints in appendix~\ref{app:constraints} and Eqs.~\eqref{eqn:QT_constraints1} and~\eqref{eqn:QT_constraints2}, we find that only seven of these terms make non-trivial contributions to the field equations; these are characterized by the constants $c_1$, $c_2$, $c_3$, $c_4$, $c_5$, $c_6$ and $c_7$.  Of these seven non-trivial theories, we know that one must correspond to quartic Lovelock gravity; i.e. there must be a choice of constants that produces the eight dimensional Euler density.  We find that this to be
\begin{align} 
\mathcal{X}_8 : \quad & c_1 =  96,  \quad c_{2} = -48, \quad  c_{3} = 96, \quad  c_{4} = -48, \quad  c_5 = -6, \quad  c_6 = 48, \quad  c_7 = -3,  
\nn\\
&c_8 = -384, \quad c_{10} = -192, \quad  c_{11} = 32, \quad  c_{13} = -192, \quad  c_{14} = 192,  \quad  c_{16} = -192  .
\end{align}
Another known term ensuring a non-trivial field equation is the selection
\begin{align}\label{qZ-1}
\mathcal{Z}_d^{(1)} : \quad 
&c_1 = 0, \quad c_2 = 8 (d-2) (860-2113 d+1959 d^2-810 d^3+102 d^4+30 d^5-11 d^6+d^7)
\nn\\
&c_3 = 0, \quad c_4 = 0, \quad  c_6 = 0 ,
\nn\\
&c_5 = -(d-2) (1108-2723 d+2639 d^2-1224 d^3+235 d^4+10 d^5-10 d^6+d^7) , 
\nn\\
&c_7 = -1292+2929 d-2741 d^2+1527 d^3-684 d^4+276 d^5-82 d^6+14 d^7-d^8,
\nn\\
&c_8 = 0, \quad c_{10}  = 0 , \quad c_{11} = 0, \quad c_{13} = 0, 
\nn\\
&c_{14} = 16 (d -2)^3 (274-389 d+183 d^2-34 d^3+2 d^4), \quad c_{16} = 0 \, . 
\end{align}
corresponding to  quartic quasi-topological gravity~\cite{Dehghani:2011vu}.  

We thus have five new quartic quasi-topological theories, which to our knowledge have not been discussed in the literature to date.  We therefore choose a simple basis for these terms:
\begin{align}
\mathcal{Z}_d^{(2)} : \quad & c_1 = 1  , \quad  \text{\small other $c_i = 0$ except those constrained in appendix~\ref{app:constraints} and Eqs.~\eqref{eqn:QT_constraints1} and \eqref{eqn:QT_constraints2}}
\nn\\
\mathcal{Z}_d^{(3)} : \quad & c_2 = 1 \, , \quad  \text{\small other $c_i = 0$ except those constrained in appendix~\ref{app:constraints} and Eqs.~\eqref{eqn:QT_constraints1} and \eqref{eqn:QT_constraints2}}
\nn\\
\mathcal{Z}_d^{(4)} : \quad & c_3 = 1 \, , \quad  \text{\small other $c_i = 0$ except those constrained in appendix~\ref{app:constraints} and Eqs.~\eqref{eqn:QT_constraints1} and \eqref{eqn:QT_constraints2}}
\nn\\
\mathcal{Z}_d^{(5)} : \quad & c_4 = 1 \, , \quad  \text{\small other $c_i = 0$ except those constrained in appendix~\ref{app:constraints} and Eqs.~\eqref{eqn:QT_constraints1} and \eqref{eqn:QT_constraints2}}
\nn\\
\mathcal{Z}_d^{(6)} : \quad & c_5 = 1 \, , \quad  \text{\small other $c_i = 0$ except those constrained in appendix~\ref{app:constraints} and Eqs.~\eqref{eqn:QT_constraints1} and \eqref{eqn:QT_constraints2}}
\label{qZ-2}
\end{align}
The resulting expressions for the Lagrangian densities of the quasi-topological theories listed above exhibit complicated dependence on the spacetime dimension.  We have included explicit expressions for these, valid in any dimension $d > 4$, in appendix~\ref{app:QT_lags}.  Each of the quasi-topological theories contributes to the field equations in dimensions $d \ge 5$, but are `quasi-topological' in $d =8$, i.e. they do not contribute to the equations of motion for eight dimensional spherically symmetric metrics, but do not correspond to a topological invariant of general metrics.  

A remark about these new quasi-topological Lagrangians is in order.   As discussed above, we have found that there are seven terms that make identical ``Lovelock-like" contributions to the field equations in the quartic case.  One of these is of course quartic Lovelock gravity, while we have found \textit{six Lagrangian densities} (only one of which~\cite{Dehghani:2011vu} was previously known) 
that would all fall into the class of theories known as quasi-topological gravity.  It is not possible to ``move between" these quasi-topological terms by adding a term proportional to the eight-dimensional Euler density:  there is no linear combination of the $\mathcal{Z}_d^{(i)}$ terms we have defined  yielding $\mathcal{X}_8$. This is in notable contrast to  the cubic case where, in five dimensions, there are two contributions to the field equations named, in the notation of~\cite{Myers:2010ru}, $\mathcal{Z}_5$ and $\mathcal{Z}'_5$.  However these densities obey the relationship~\cite{Myers:2010ru}
\be 
\mathcal{X}_6 = 4 \mathcal{Z}'_5 - 8 \mathcal{Z}_5  
\ee
and since the six dimensional Euler density identically vanishes in five dimensions for any metric, it follows that there are not really two independent theories.  \textit{Cubic quasi-topological gravity is unique}.  The fact that in the quartic case
\be 
\mathcal{X}_8 \neq \sum_{i=1}^6 c_i \mathcal{Z}_d^{(i)}
\ee  
for any choice of the coefficients $c_i$ means that each of these theories are distinct for general metrics.  However, as mentioned at the beginning of this section, there is a sense in which these theories are degenerate.  Under the constraint of spherical symmetry, there exist invariants that vanish for any spherically symmetric metric~\cite{Deser:2005pc}.  In fact, the combination
\be 
I^{(ij)} = \frac{\hat{\mu}_{(i)}}{\mu_{(i)}} \mathcal{Z}^{(i)} - \frac{\hat{\mu}_{(j)}}{\mu_{(j)}} \mathcal{Z}^{(j)}
\ee
will always be such a term [the quantities with the hats are defined below in Eq.~\eqref{eqn:normalized_couplings}].  Thus, in spherical symmetry, there is a ``unique" quasi-topological theory in the sense that each of the $\mathcal{Z}_d^{(i)}$ terms makes the same contribution to the field equations and are related to one another by the addition of a term that vanishes on \textit{spherically symmetric} metrics.  We emphasize, however, that these theories are ultimately distinct because they will each yield different dynamics when   spherical symmetry
is not imposed.

The  quartic quasi-topological term \eqref{qZ-1}  
  was also claimed to be unique;  however this does not appear to be the case, at least in the sense originally described \cite{Dehghani:2011vu}.  That theory is unique only in the sense described above: terms  vanishing under the constraint of spherical symmetry can be added to the action without altering the field equations.  However, away from spherical symmetry these will be distinct theories, even in less than eight dimensions. 

Many of these considerations are nicely discussed in the recent paper~\cite{Cisterna:2017umf}, where a quintic quasi-topological theory was presented.  It would be interesting to know how many theories contribute to the field equations in that case.   More ambitiously, it would be interesting to know more general criteria that  allows one to construct the quasi-topological Lagrangians in general, a problem that remains open.

\subsubsection{Generalized quasi-topological terms}

We now move on to consider generalized quasi-topological terms.  Of the 13 free parameters remaining under the restrictions imposed by the constraints in appendix~\ref{app:constraints} and Eqs.~\eqref{eqn:QT_constraints1} and \eqref{eqn:QT_constraints2}, the Lovelock and six quasi-topological terms comprise the only seven non-trivial theories.  The remaining six terms do not contribute to the field equations of a VSSS ansatz.  Here we do not explicitly present the Lagrangians for these terms, but rather simply indicate choices of constants by which they are produced.  We make the following choices:
\begin{align} 
\mathcal{C}^{(1)}_d  : \quad & c_8 = 1 \, , \quad  \text{\small other $c_i = 0$ except those constrained in appendix~\ref{app:constraints} and Eqs.~\eqref{eqn:QT_constraints1} and \eqref{eqn:QT_constraints2}}
\nn\\
\mathcal{C}^{(2)}_d  : \quad & c_{10} = 1 \, , \quad  \text{\small other $c_i = 0$ except those constrained in appendix~\ref{app:constraints} and Eqs.~\eqref{eqn:QT_constraints1} and \eqref{eqn:QT_constraints2}}
\nn\\
\mathcal{C}^{(3)}_d  : \quad & c_{11} = 1 \, , \quad  \text{\small other $c_i = 0$ except those constrained in appendix~\ref{app:constraints} and Eqs.~\eqref{eqn:QT_constraints1} and \eqref{eqn:QT_constraints2}}
\nn\\
\mathcal{C}^{(4)}_d  : \quad & c_{13} = 1 \, , \quad  \text{\small other $c_i = 0$ except those constrained in appendix~\ref{app:constraints} and Eqs.~\eqref{eqn:QT_constraints1} and \eqref{eqn:QT_constraints2}}
\nn\\
\mathcal{C}^{(5)}_d  : \quad & c_{14} = 1 \, , \quad  \text{\small other $c_i = 0$ except those constrained in appendix~\ref{app:constraints} and Eqs.~\eqref{eqn:QT_constraints1} and \eqref{eqn:QT_constraints2}}
\nn\\
\mathcal{C}^{(6)}_d  : \quad & c_{16} = 1 \, , \quad  \text{\small other $c_i = 0$ except those constrained in appendix~\ref{app:constraints} and Eqs.~\eqref{eqn:QT_constraints1} and \eqref{eqn:QT_constraints2}}
\label{C-theories}
\end{align}
These remaining six terms yield vanishing contributions to the field equations when $N = 1$ is permitted (e.g. in vacuum or for electromagnetic matter), but would make non-vanishing contributions in the presence of more general matter distributions. 

We shall now relax the additional constraints imposed in Eqs.~\eqref{eqn:QT_constraints1} and \eqref{eqn:QT_constraints2} in order to obtain the full family of theories satisfying the constraint~\eqref{eqn:feq_condition}.  These four distinct new theories -- the quartet --  have a
field equation that is a total derivative of a quantity that is a polynomial in both $f(r)$ and its first two derivatives.
We make the following selections:
\begin{align}
\label{eqn:picking_GQT}
\mathcal{S}_d^{(1)} : \quad  c_1 &= 1 \, ,
\nn\\
c_9 &= -{\frac {2{d}^{6}-23{d}^{5}+106{d}^{4}-292{d}^{3}+588{d}^{2}
-709d+320}{d \left( d-3 \right)  \left( 3{d}^{2}-18d+19 \right) 
 \left( {d}^{2}-6d+11 \right) }} \, ,
\nn\\
c_{15} &= {\frac {1}{ \left( d-3 \right) ^{2} \left( {d}^{3}-9{d}^{2}+26d-22
 \right) d \left( 3{d}^{2}-18d+19 \right)  \left( {d}^{2}-6d+11
 \right) }} \times
 \nn\\
&\times \big[{d}^{10}-20{d}^{9}+188{d}^{8}-1211{d}^{7}+6287{d}^{6}-
25778{d}^{5}+75674{d}^{4}
\nn\\
&-146251{d}^{3}+172418{d}^{2}-110076
d+28160 \big]  \, ,
\nn\\
&\text{All other $c_i = 0$ except those constrained in appendix~\ref{app:constraints}} \, ,
\nn\\
\mathcal{S}_{d}^{(2)} : \quad c_2 &= 1 \, ,
\nn\\
 c_9 &= -{\frac {2 \left( 2{d}^{6}-24{d}^{5}+103{d}^{4}-161
{d}^{3}-67{d}^{2}+409d-274 \right) }{d \left( d-3 \right) 
 \left( 3{d}^{2}-18d+19 \right)  \left( {d}^{2}-6d+11 \right) }} \, ,
\nn\\
c_{15} &= {\frac {2 }{ \left( d-3 \right) ^{2}
 \left( {d}^{3}-9{d}^{2}+26d-22 \right) d \left( 3{d}^{2}-18d+
19 \right)  \left( {d}^{2}-6d+11 \right) }} \times
\nn\\
&\times\big[  {d}^{10}-16{d}^{9}+55{d}^{8}+601{d}
^{7}-7258{d}^{6}+35933{d}^{5}-102275{d}^{4}
\nn\\
&+177665{d}^{3}-
184591{d}^{2}+104237d-24112 \big] \, ,
\nn\\
&\text{All other $c_i = 0$ except those constrained in appendix~\ref{app:constraints}} \, ,
\nn\\
\mathcal{S}_d^{(3)} : \quad c_4 &= 1 \, ,
\nn\\ 
c_{9} &= -{\frac {4 (2{d}^{6}-25{d}^{5}+112{d}^{4}-185{d}^{3}-70{d}^{
2}+494d-340)}{d \left( d-3 \right)  \left( 3{d}^{2}-18d+19
 \right)  \left( {d}^{2}-6d+11 \right) }} \, ,
\nn\\
c_{15} &= {\frac {4}{ \left( d-3 \right) ^{2} \left( {d}^{3}-9{d}^{2}+26d-22
 \right) d \left( 3{d}^{2}-18d+19 \right)  \left( {d}^{2}-6d+11
 \right) }} \times
 \nn\\
 &\times \big[{d}^{10}-18{d}^{9}+99{d}^{8}+193{d}^{7}-5212{d}^{6}
+30115{d}^{5}-93864{d}^{4}
\nn\\
&+175930{d}^{3} -196892{d}^{2}+120000
d-29920 \big] \, ,
\nn\\
&\text{All other $c_i = 0$ except those constrained in appendix~\ref{app:constraints}} \, ,
\nn\\
\mathcal{S}_d^{(4)} : \quad c_{5} &= 1 \, ,
\nn\\
c_{9} &= -{\frac {8(2{d}^{6}-25{d}^{5}+112{d}^{4}-185{d}^{3}-70{d}^{
2}+494d-340)}{d \left( d-3 \right)  \left( 3{d}^{2}-18d+19
 \right)  \left( {d}^{2}-6d+11 \right) }} \, ,
\nn\\
c_{15} &= {\frac {8}{ \left( d-3 \right) ^{2} \left( {d}^{3}-9{d}^{2}+26d-22
 \right) d \left( 3{d}^{2}-18d+19 \right)  \left( {d}^{2}-6d+11
 \right) }} \times
 \nn\\
 &\times \big[{d}^{10}-18{d}^{9}+99{d}^{8}+193{d}^{7}-5212{d}^{6}
+30115{d}^{5}-93864{d}^{4}
\nn\\
&+175930{d}^{3}-196892{d}^{2}+120000
d-29920 \big]
\nn\\
&\text{All other $c_i = 0$ except those constrained in appendix~\ref{app:constraints}} \, .
\end{align} 
We have chosen these constants to render the field equations in general dimensions as simple as possible. Our choices have been further motivated by the four dimensional case, which will be presented in the following subsection. The explicit Lagrangian densities that result for these terms are presented in appendix~\ref{app:GQT_lags}. 

Although we have made many different attempts, it does not seem possible to select additional constraints such that the reduced Lagrangian of these generalized quasi-topological theories takes the form,
\be 
L_{N,f} = N F_{0} + N' F_{1} + N''F_{2} 
\ee
where $F_{i}$ are functions of $f$ and its derivatives and the primes denote differentiation with respect to $r$.  In other words, it does not seem possible to eliminate terms that are higher order in the derivatives of $N$ (e.g. $N'^2/N$, etc.) without also eliminating the theory.  This adds further support to Conjecture 2 made in~\cite{Bueno:2017sui} since we also find that the field equations for these theories are not algebraic.

We have now listed a basis for all 17 theories which satisfy condition~\eqref{eqn:feq_condition} at the quartic level.  We are now able to write down  the explicit action for the full theory in five and higher dimensions. This takes the form
\begin{align} 
\mathcal{I} = \frac{1}{16 \pi G} \int d^d x \sqrt{-g} \bigg[&-2 \Lambda + R + \alpha_2 \mathcal{X}_4+  \alpha_3 \mathcal{X}_6 + \mu \mathcal{Z}_d -  \lambda \mathcal{S}_d + \alpha_4 \mathcal{X}_8 
\nn\\
&+ \sum_{i=1}^6 \hat{\mu}_{(i)} \mathcal{Z}_d^{(i)} - \sum_{i=1}^{4} \hat{\lambda}_{(i)} \mathcal{S}_d^{(i)} + \sum_{i=1}^{6} \gamma_{(i)} \mathcal{C}_d^{(i)}   \bigg]
\end{align} 
For any situation in which the stress energy tensor satisfies $T_{t}^t = T_{r}^r$ (including the vacuum) the $\mathcal{C}_d^{(i)}$ terms will make no contribution to the field equations: their contributions to the generalized Einstein tensor all contain derivatives of $N$.  For this reason, we shall not include these terms in our action in any of the discussion to follow.

In the above, we have made the following rescalings of the coupling constants to simplify the resulting field equations:
\begin{align}
\label{eqn:normalized_couplings}
\hat{\mu}_{(1)} & = {\frac {1}{ \left( d-1 \right)    \left( d-2 \right) \left( d-3 \right)^{2}  \left( d-4
 \right) \left( d-8 \right)   P}}
 \mu_{(1)} \, ,
\nn\\
\hat{\mu}_{(2)} & = {\frac {24(3d-4)}{ \left( d-3 \right) \left( d-
8 \right)  \left( 28{d}^{3}-173d+160+{d}^{5}+18{d}^{2}-10{d}^{
4} \right) }} \mu_{(2)}\, ,
\nn\\
\hat{\mu}_{(3)} & = {\frac {12(3d-4)}{ \left( d-3 \right)  \left( d-
8 \right)  \left( -12{d}^{4}+61{d}^{3}+242d-167{d}^{2}+{d}^{5}
-137 \right) }} \mu_{(3)}\, ,
\nn\\
\hat{\mu}_{(4)} & = {\frac {12(3d-4)}{ \left( d-3 \right)   \left( d-
8 \right)  \left( d-4 \right)  \left( {d}^{3}-10{d}^{2}+31d-26
 \right) }} \mu_{(4)}\, ,
\nn\\
\hat{\mu}_{(5)} & = {\frac {6(3d-4)}{ \left( d-3 \right)  \left( d-8
 \right)  \left( {d}^{5}+79{d}^{3}-14{d}^{4}+316d-170-224{d}^{
2} \right) }} \mu_{(5)}\, ,
\nn\\
\hat{\mu}_{(6)} & =
{\frac {3(3d-4)}{ \left( d-3 \right)   \left( d-8
 \right)  \left( {d}^{5}+79{d}^{3}-14{d}^{4}+316d-170-224{d}^{
2} \right) }} \mu_{(6)}\,,
\nn\\
\hat{\lambda}_{(1)} &= {\frac { d\left( 3{d}^{3}-27{d}^{2}+73d-57 \right)}{   \left( 2{d}^{5}-20{d}^{4}+56{d}^{3}+36{d}^{2}-346d
+320 \right) }} \lambda_{(1)} \, ,
\nn\\
\hat{\lambda}_{(2)} &= {\frac { \left( 3{d}^{3}-27{d}^{2}+73\,d-57 \right) d}{ 4   \left( {d}^{5}-12{d}^{4}+61{d}^{3}-167{d}^{2}+242\,d-
137 \right) }} \lambda_{(2)}\,,
\nn\\
\hat{\lambda}_{(3)} &= {\frac { \left( 3{d}^{3}-27{d}^{2}+73\,d-57 \right) d}{8
 \left( {d}^{5}-14{d}^{4}+79{d}^{3}-224{d}^{2}+316\,d-170
 \right)   }} \lambda_{(3)} \, ,
\nn\\
\hat{\lambda}_{(4)} &= {\frac { \left( 3{d}^{3}-27{d}^{2}+73\,d-57 \right) d}{16  \left( {d}^{5}-14{d}^{4}+79{d}^{3}-224{d}^{
2}+316\,d-170 \right) }} \lambda_{(4)}\, .
\end{align}
In the first term above we have defined
\begin{align}
P =& \left( {d}^{5}-20\,{d}^{4}+142\,{d}^{3}-472{d}^{2}+743\,d-
436 \right)
\end{align}

This concludes our discussion of the theories in dimensions larger than four.  We now turn to a discussion of the four dimensional case.

\subsection{The case for four dimensions}

As was the case in the cubic version of generalized quasi-topological gravity, the four dimensional case is somewhat special, with only seven constraints as opposed to nine (see previous subsection).  We find that the most general four dimensional theory satisfying~\eqref{eqn:feq_condition} is given by placing the following seven constraints on the quartic terms in the action:
\begin{align}
\label{eqn:4d_constraints}
c_{12} &= -\frac{19}{60} c_1- \frac{1}{2}c_2-\frac{1}{12}c_3-\frac{4}{5}c_4-\frac{8}{5}c_5-\frac{14}{15}c_6-\frac{56}{15}c_7-\frac{1}{8}c_8-\frac{1}{4}c_9-\frac{1}{2}c_{11}  ,
\nn\\
c_{17} &= -\frac{23}{30}c_1-\frac{4}{3}c_2-\frac{1}{12}c_3-\frac{11}{5}c_4-\frac{22}{5}c_5-\frac{41}{15}c_6-\frac{28}{5}c_7-\frac{1}{24}c_8-c_9-\frac{11}{12}c_{10}
\nn\\
&-\frac{1}{6}c_{13}-\frac{1}{3}c_{14}-\frac{1}{12}c_{15}-\frac{1}{4}c_{16}\,,
\nn\\
c_{19} &= \frac{11}{30}c_1+\frac{7}{12}c_2+\frac{1}{12}c_3+\frac{9}{10}c_4+\frac{9}{5}c_5+\frac{17}{15}c_6+\frac{16}{5}c_7+\frac{5}{48}c_8+\frac{1}{4}c_9+\frac{1}{6}c_{10}
\nn\\
&+\frac{3}{8}c_{11}-\frac{1}{48}c_{13}+\frac{1}{12}c_{14}-\frac{1}{24}c_{15}-\frac{1}{4}c_{18} \,,
\nn\\
c_{20} &= \frac{36}{5} c_1+\frac{32}{3} c_2+2 c_3+\frac{72}{5} c4+\frac{144}{5} c_5+ \frac{104}{5} c_6+ \frac{1088}{15} c_7+ \frac{7}{3} c_8+ \frac{4}{3} c_{10}+8 c_{11}
\nn\\
&- \frac{8}{3} c_{13}+ \frac{2}{3} c_{14}- \frac{10}{3} c_{15}-2 c_{16}-4 c_{18} +2 c_{24}+8 c_{25}+32 c_{26} \, ,
\nn\\
c_{21} &= - \frac{5}{3} c_1- \frac{8}{3} c_2-\frac{1}{3} c_3-4 c_4-8 c_5- \frac{16}{3} c_6-\frac{32}{3} c_7-\frac{1}{3} c_8-c_9-\frac{4}{3} c_{10}-c_{11}
\nn\\
&+\frac{1}{6} c_{13}- \frac{2}{3} c_{14}+\frac{1}{3} c_{15}-c_{24}-4 c_{25}-16 c_{26} \, ,
\nn\\
c_{22} &= - \frac{7}{3} c_1- \frac{10}{3} c_2- \frac{2}{3} c_3-4 c_4-8 c_5-\frac{20}{3} c_6- \frac{64}{3} c_7- \frac{2}{3} c_8- \frac{2}{3} c_{10}-2 c_{11}
\nn\\
&+ \frac{1}{3} c_{13}- \frac{1}{3} c_{14}+ \frac{2}{3} c_{15}-2 c_{24}+16 c_{26} \, ,
\nn\\
c_{23} &= \frac{1}{15} c_1+ \frac{1}{3} c_2-\frac{1}{6} c_3+ \frac{4}{5} c_4+ \frac{8}{5} c_5+ \frac{14}{15} c_6- \frac{28}{5} c_7- \frac{1}{3} c_8+c_9+ \frac{7}{6} c_{10}- \frac{3}{2} c_{11}
\nn\\
&+ \frac{5}{12} c_{13} + \frac{1}{3} c_{14}+ \frac{1}{3} c_{15} + \frac{1}{2} c_{16} +c_{18}-2 c_{25}-12 c_{26} \, .
\end{align} 
Thus, one is left with a 19 parameter family of quartic densities whose solutions are of the form Eq.~\eqref{eqn:metricAnsatz}
with $N = \,$constant.  We shall now discuss a useful basis for these theories.  

In general, only the six terms corresponding to $c_1$, $c_2$, $c_4$, $c_5$, $c_6$ and $c_7$ make nonzero contribute to field equations in the context of VSSS metrics. Furthermore, each of these six terms make the same contributions to the field equations, up to overall constants.  These six terms provide the quartic generalizations of the cubic $\mathcal{S}_4$ term in four dimensions. The remaining 13 terms do not contribute to the field equations of a VSSS ansatz, or in any case where the stress energy tensor satisfies $T_t^t = T_r^r$. Our focus here will be to present the six non-vanishing contributions.

In the previous subsection we presented four Lagrangian densities, $\mathcal{S}_d^{(i)}$ with $i = 1,2,3,4$.  These terms account for four of the six contributions in four dimensions, upon setting $d=4$ in the expressions presented in appendix~\ref{app:GQT_lags}.  The two additional non-trivial contributions can be obtained by the following selection of free parameters.
\begin{align}
\mathcal{S}_4^{(5)} : \quad &c_6 = 1 \, , \quad c_9 = -\frac{56}{15}
\nn\\
&\text{All other $c_i = 0$ except those constrained in Eq.~\eqref{eqn:4d_constraints}} \, ,
\nn\\
\mathcal{S}_4^{(6)} : \quad &c_7 = 1 \, , \quad c_9 = - \frac{224}{15}
\, ,
\nn\\
&\text{All other $c_i = 0$ except those constrained in Eq.~\eqref{eqn:4d_constraints}} \, .
\label{eqn:4d_GQT}
\end{align}
We have presented explicit forms for these expressions in appendix~\ref{app:GQT_lags}.  In addition to the six non-trivial terms, there are 13 terms that are the four dimensional analogs of the $\mathcal{C}_d^{(i)}$ terms.  We do not present full expressions for these terms here since they have no effect on the field equations in the situations we are interested in.  A simple basis for these terms is obtained simply taking $c_i =1$ and all other $c_j = 0$ (except those which are constrained) for each of the constants that have not been fixed by the above considerations.   

We note again that the imposition of spherical symmetry yields a degeneracy amongst these theories: they differ by terms  that vanish for a spherically symmetric metric.  However this degeneracy is lifted if spherical symmetry is relaxed and so
 the theories are ultimately distinct.

 The action for the non-trivial contributions to the field equations in four dimensions reads
\begin{align} 
\mathcal{I} = \frac{1}{16 \pi G} \int d^4 x \sqrt{-g} \bigg[&-2 \Lambda + R -  \lambda \mathcal{S}_4  - \sum_{i=1}^{6} \hat{\lambda}_{(i)} \mathcal{S}_4^{(i)} \bigg]
\end{align} 
where
\be
\label{eqn:normalized_4d_coupling} 
\hat{\lambda}_{(5)} = -\frac{5}{24} \lambda_{(5)} \, , \quad \hat{\lambda}_{(6)} = -\frac{5}{96} \lambda_{(6)} 
\ee
with all other $\hat{\lambda}_{(i)}$ as defined in Eq.~\eqref{eqn:normalized_couplings} with $d = 4$. These choices of normalization have been made to simplify the form of the field equations.

\section{Linear theory and vacua}
\label{sec:linear}

In this section we provide a brief discussion of the linearized equations of motion for the theories presented in the section~\ref{sec:construction}.  As  conjectured in~\cite{Hennigar:2017ego} and then demonstrated in~\cite{Bueno:2017sui}, a theory satisfying Eq.~\eqref{eqn:feq_condition} must necessarily have linearized equations of motion that agree with the linearized Einstein equations on a maximally symmetric background, up to an overall constant.  The only caveat being that, in this theory, the metric~\eqref{eqn:metricAnsatz} describes the gravitational field outside of a spherically symmetric mass distribution. Thus, this section provides a useful check of the correctness  of the theories, and the results may be useful in future studies of these theories.

In what follows we will follow closely reference~\cite{Bueno:2016ypa}, adopting the conventions therein.  We consider a perturbation $h_{ab}$ away from a maximally symmetric spacetime  $\bar{g}_{ab}$ such that,
\be 
g_{ab} = \bar{g}_{ab} + h_{ab} \, .
\ee
The curvature of the maximally symmetric background is given by,
\be 
\bar{R}_{abcd} = 2 K \bar{g}_{a[c}\bar{g}_{d] b} \, . 
\ee
The linearized equations of motion for $h_{ab}$ are then given by~\cite{Bueno:2016ypa},
\begin{align} 
\frac{1}{2} \mathcal{E}^L_{ab} =& \big[e - 2 K\left( a (d-1) + c \right) + (2a + c) \bar{\square} \big] G_{ab}^L
+ [ a + 2b + c ] \big[\bar{g}_{ab} \bar{\square} - \bar{\nabla}_a\bar{\nabla}_b  \big] R^L 
\nn\\
& - K \big[a(d-3) - 2b(d-1) - c \big]\bar{g}_{ab} R^L = \frac{1}{4} T^L_{ab}
\end{align}
where $a$, $b$, $c$ and $e$ are a convenient choice of parameters based on the linearization procedure; see~\cite{Bueno:2016ypa} for further details.  In the above, all quantities with a bar correspond to those defined for the background metric, $\bar{g}_{ab}$, while
\begin{align}
G_{ab}^L &= R_{ab}^L - \frac{1}{2} \bar{g}_{ab} R^L - (d-1)K h_{ab} \, ,
\nn\\
R_{ab}^L &= \frac{1}{2} \big[\bar{\nabla}_a \bar{\nabla}_c h_b{}^c + \bar{\nabla}_b \bar{\nabla}_c h_a{}^c - \bar{\square} h_{ab} -  \bar{\nabla}_a \bar{\nabla}_b h \big] + dK h_{ab} - K h \bar{g}_{ab}
\nn\\
R^L &= \bar{\nabla}^a \bar{\nabla}^b h_{ab} - \bar{\square} h - (d-1) K h
\end{align}
where $h = \bar{g}^{ab}h_{ab}$.  The additional scalar and massive graviton modes will be absent provided $2a + c = 0$ and $4b + c = 0$~\cite{Bueno:2016ypa}.   In other words, these terms will be absent provided the linearized equations are proportional to the linearized Einstein tensor (plus cosmological term) on the same background.  Let us now explicitly present the linearized equations for the theories we have constructed.  Specifically, we consider the theory
\begin{align} 
\label{eqn:QT_theory}
\mathcal{I} = \frac{1}{16 \pi G} \int d^d x \sqrt{-g} \bigg[ -2 \Lambda + R  - \sum_{i=1}^{4} \hat{\lambda}_{(i)} \mathcal{S}_d^{(i)} + \sum_{i=1}^6 \hat{\mu}_{(i)} \mathcal{Z}_d^{(i)} \bigg] \, ,
\end{align}
 which includes all of the non-trivial contributions at the quartic level, except for the Lovelock term, which has been thoroughly studied.  The results can be easily extended to cases with additional terms appearing in the action by simply adding the contributions arising from these terms to the relevant equations below.

We will define, for convenience, the following constants:
\be 
\mu \coloneqq \sum_{i=1}^{6} \mu_{(i)} \, , \quad \lambda \coloneqq \sum_{i=1}^{4 + 2\delta_{d,4}} \lambda_{(i)} \, .
\ee
Then it is a matter of calculation to show that the linearized equations are given by,
\begin{align}
\mathcal{E}_{ab}^L = \frac{1}{2}\bigg[1 + 4 \left( \mu + \frac{(d-8)}{3}   \lambda \right) K^3   \bigg] G_{ab}^L
\end{align}
Note that in the above, it is the couplings without hats that appear; the definitions made in Eq.~\eqref{eqn:normalized_couplings} significantly simplify the form of the linearized equations.   As expected, we see the result is  proportional to the Einstein tensor linearized on the same background.

In four dimensions, the additional terms $\mathcal{S}_4^{(5)}$ and $\mathcal{S}_4^{(6)}$ also contribute, while the quasi-topological terms make no contribution.  The linearized field equations then become
\begin{align}
\mathcal{E}_{ab}^L = \frac{1}{2}\bigg[1  - \frac{16}{3}\lambda K^3   \bigg]  G_{ab}^L \, , \quad \text{ in $d = 4$} 
\end{align}
where the sum defining $\lambda$ now runs over all six couplings, $\lambda_{(i)}$. 

The full field equations will relate the curvature of the background, $K$, to the length scale introduced by the cosmological constant, $\Lambda$.  This dependence can be obtained by evaluating the field equations (see next section) on the maximally symmetric background.  One finds that the following relationship must hold,
\be 
- \frac{2 \Lambda}{(d-1)(d-2)} + K + \left( \mu + \frac{(d-8)}{3} \lambda \right) K^4 = 0 \, ,
\ee 
with $\mu$ and $\lambda$ defined by the sums above. Note that when the higher curvature terms are switched off, the cosmological constant uniquely determines the curvature of the maximally symmetric solutions of the theory.  However, when the higher curvature terms are present there will generically be multiple solutions for $K$: four in this quartic theory. In general, only a single one of these solutions will have a smooth limit to the vacuum of Einstein gravity upon sending $\mu, \lambda \to 0$. 

In order to ensure the proper coupling to matter, the prefactor appearing in front of $G_{ab}^L$ in the linearized equations must have the same sign as in Einstein gravity.  If this were not the case, then the graviton would be a ghost.  For the theory discussed here, this requirement demands,
\be 
1 + 4 \left( \mu + \frac{(d-8)}{3}   \lambda \right) K^3   > 0 \, .
\ee 
This condition must be satisfied by any physically reasonable solution to the equations of motion.  

We close this section by noting that, in a $d > 4$ theory that contains both the quasi-topological and generalized quasi-topological terms, the value
\be 
\mu = - \frac{(d-8)}{3} \lambda
\ee
seems to be special.  When the couplings are constrained in this way, the theory has a unique vacuum coinciding with the Einstein gravity vacuum.  Further, the above inequality for the absence of ghosts is trivially satisfied.  It would be interesting to see if there are any additional interesting properties associated with this value of $\mu$.

\section{Nonlinear field equations in spherical symmetry}
\label{sec:fieldeqns}

Here we present the field equations that are derived from the actions presented in section~\ref{sec:construction}.  We consider first the theory defined in dimensions larger than four, and then close with the four dimensional case.

\subsection{The field equations in dimensions larger than four}

\subsubsection{Quasi-topological theories}

We consider first the field equations for the quasi-topological gravities constructed in section~\ref{sec:construction}.  The field equations of Lovelock gravity are well known and we do not discuss them here.  We consider the following action,
\begin{align} 
\label{eqn:QT_theory}
\mathcal{I} = \frac{1}{16 \pi G} \int d^d x \sqrt{-g} \bigg[ \frac{(d-1)(d-2)}{\ell^2} + R  + \sum_{i=1}^6 \hat{\mu}_{(i)} \mathcal{Z}_d^{(i)} \bigg]
\end{align} 
where the $\hat{\mu}_{(i)}$ terms are as in Eq.~\eqref{eqn:normalized_couplings}.  A spherically symmetric metric~\eqref{eqn:metricAnsatz} satisfies the field equations $F' = 0$ with $N = 1$ and
\be\label{QT-eq} 
F = (d-2)r^{d-1}\left[\frac{1}{\ell^2} - \psi +  \sum_{i=1}^{6} \mu_{(i)} \psi^4 \right]  
\ee
where we have defined
\be 
\psi = \frac{f - k}{r^2}\, .
\ee
Equation \eqref{QT-eq} can be easily integrated revealing that 
 $f$ is determined by the following algebraic relationship,
\be 
(d-2)r^{d-1}\left[\frac{1}{\ell^2} - \psi +  \sum_{i=1}^{6} \mu_{(i)} \psi^4 \right] = m
\ee
where $m$ is an integration constant which is related to the mass.  Note that in passing from the action to the field equations, the hats have been removed from the $\mu$'s.  It was for this simplification that the $\hat{\mu}_{(i)}$ terms were defined as in Eq.~\eqref{eqn:normalized_couplings}.  Note that these equations only hold for $d > 4$ but $ d\neq 8$: in eight dimensions, the quasi-topological terms are trivial.

\subsubsection{Generalized quasi-topological theories} 

We next present the field equations for the four non-trivial generalized quasi-topological terms that were presented in section~\ref{sec:construction}.  The field equations for these theories are not algebraic, but rather, in vacuum, integrate to a second order differential equation that the metric function $f$ must satisfy.  

We consider now the following action,
\begin{align} 
\label{eqn:GQT_theory}
\mathcal{I} = \frac{1}{16 \pi G} \int d^d x \sqrt{-g} \bigg[ \frac{(d-1)(d-2)}{\ell^2} + R  - \sum_{i=1}^{4} \hat{\lambda}_{(i)} \mathcal{S}_d^{(i)} \bigg]
\end{align}
where the $\hat{\lambda}_{(i)}$ terms are as defined in Eq.~\eqref{eqn:normalized_couplings}.  The field equations of this theory can be written in the following simple form,
\be 
F' = 0
\ee
where 
\be 
F = (d-2)r^{d-3}\left[k - f + \frac{r^2}{\ell^2}  \right] + (d-2)\left( \sum_{i=1}^4 \lambda_{(i)} \right) F_{\mathcal{S}_d}
\ee
and the $F_{\mathcal{S}_d}$ represents the contribution from each $\mathcal{S}_d^{(i)}$ to the field equations, which is the same for each term $\mathcal{S}_d^{(i)}$ due to the choices  made in Eq.~\eqref{eqn:picking_GQT}.  Note that, once again, in passing from the action to the field equations, the hats have been removed from the $\lambda$'s.  It was for this simplification that they were normalized in this way in Eq.~\eqref{eqn:normalized_couplings}.  Explicitly, the contribution made to the field equations from each $\mathcal{S}_d^{(i)}$ is given by
\begin{align}
\label{eqn:GQT_eom}
F_{\mathcal{S}_d} &=   \left( k-f \right) \left[   \left( d-4
 \right) f \left( k-f \right)  f''+{f'}^{2} \left(  \left({d}^{2}-\frac{23}{2} d + 32
 \right) f- \frac{1}{2}\,k \left( d-4 \right)  \right)  \right] {r}^{d-7}
 \nn\\
 &+
  2\,f f' f'' \left(  \left( k-f \right)  \left( d-5 \right) {r}^{d-6}+
\frac{f'}{8} \left( 3d- 16 \right) {r}^{d-5} \right)  
\nn\\
& +f f'
 \left( k-f \right) ^{2} \left( d-4 \right)  \left( d-7 \right) {r}^{d-8}+\frac{f'^3}{12} \bigg[  \left(  \left( 3d- 16 \right) f-8 k
 \right)  \left( d-5 \right) {r}^{d-6}
  \nn\\
& 
- 3 \frac{f'}{4}\left( 3d-16 
 \right) {r}^{d-5} \bigg] \, ,    
\end{align}
where $f = f(r)$ and a prime denotes a derivative with respect to $r$.
  
\subsection{The field equations in four dimensions}

In four dimensions, the only non-trivial contributions to the field equations come from the generalized quasi-topological terms.  Considering the action
\begin{align} 
\mathcal{I} = \frac{1}{16 \pi G} \int d^4 x \sqrt{-g} \bigg[&-2 \Lambda + R  - \sum_{i=1}^{6} \hat{\lambda}_{(i)} \mathcal{S}_4^{(i)} \bigg]
\end{align} 
with the $\hat{\lambda}_{(i)}$ terms defined in Eqs.~\eqref{eqn:normalized_couplings} and~\eqref{eqn:normalized_4d_coupling}, the field equations read
\be 
F' = 0
\ee
where 
\be 
F = 2r\left[k - f + \frac{r^2}{\ell^2}  \right] + 2 \left( \sum_{i=1}^6 \lambda_{(i)} \right) F_{\mathcal{S}_4}
\ee
and  $F_{\mathcal{S}_4}$ is given by the same expression as in Eq.~\eqref{eqn:GQT_eom} evaluated in $d=4$.  Explicitly, this takes the relatively simple form,
\begin{align}
F_{\mathcal{S}_4} 
&= 2\frac{f f' f''}{r^2} \left(f - \frac{1}{2} r f' - k \right) + \frac{f'^4}{4 r} + \frac{f'^3}{3 r^2} \left(f + 2k \right) + 2\frac{ff'^2}{r^3} (k-f) \, .
\end{align}
Having presented the field equations for the new quartic theories, we now move on to a discussion of their black hole solutions.

\section{Black holes}
\label{sec:blackholes}

\subsection{Black hole entropy}

We begin with a discussion of the black hole entropy for the various theories considered so far in this work.  In a higher curvature theory of gravity, the Bekenstein-Hawking entropy is modified by additional terms.  These terms can be calculated using the Iyer-Wald prescription~\cite{Wald:1993nt, Iyer:1994ys} where the entropy is given by,
\be 
S = -2 \pi \oint d^{d-2} x \sqrt{\gamma} \,  E^{abcd} \hat{\varepsilon}_{ab} \hat{\varepsilon}_{cd} 
\ee
where
\be 
E^{abcd} = \frac{\partial \mathcal{L}}{\partial R_{abcd}}
\ee 
and $\hat{\varepsilon}_{ab}$ is the binormal to the horizon, normalized to satisfy $\hat{\varepsilon}_{ab}\hat{\varepsilon}^{ab} = -2$. The integral is evaluated on the horizon of the black hole, which has induced metric $\gamma_{ab}$ and $\gamma = \det \gamma_{ab}$.  In the following we will present the entropy densities,
\be 
s = \frac{S}{\omega_{(d-2), k}}
\ee
where $\omega_{(d-2), k}$ is the volume of the submanifold with line element $d\Sigma_{(d-2), k}^2$.

Considering first the quartic quasi-topological theories, we find that each $\mathcal{Z}^{(i)}_d$ makes the same contribution to the Wald entropy.  For the theory written down in Eq.~\eqref{eqn:QT_theory}, this contribution to the entropy reads,
\be 
s_4^{(i)} = \frac{(d-2)}{(d-8)} \frac{k^3}{r_+^6} \mu_{(i)} r_+^{d-2} \, , \quad \text{ for  $\mathcal{Z}^{(i)}_d$}
\ee 
giving
\be 
s = \frac{r_+^{d-2}}{4} \left[1 +  \frac{(d-2)}{(d-8)} \frac{k^3}{r_+^6} \left( \sum_{i=1}^{6}  \mu_{(i)} \right) \right]
\ee
for the theory~\eqref{eqn:QT_theory}.  For a theory containing additional terms, the above entropy density is simply modified by the addition of the entropy densities corresponding to the extra terms.

In the case of the generalized quasi-topological terms, $\mathcal{S}_d^{(i)}$, due to our choices of constraints and the normalization of $\hat{\lambda}_{(i)}$ in Eq.~\eqref{eqn:normalized_couplings}, we find the same contribution to the entropy from each $\mathcal{S}_d^{(i)}$.  For the particular case of the theory presented in Eq.~\eqref{eqn:GQT_theory} this contribution reads,
\be 
s_4^{(i)} =   -\frac{\pi (d-2) T}{ r_+^3} r_+^{d-2} \left[  \frac{(d-4)k^2}{r_+^2}  +  \frac{ 4 \pi (d-5) k T}{r_+} +  \frac{(4 \pi)^2}{12} (3d - 16) T^2 \right] \lambda_{(i)}
\ee 
giving for the theory in Eq.~\eqref{eqn:GQT_theory},
\begin{align}
s &= \frac{r_+^{d-2}}{4} \bigg\{1 - \sum_{i=1}^{4} \frac{4 \pi (d-2) T}{r_+^3}  \left[  \frac{(d-4)k^2}{r_+^2}  +  \frac{ 4 \pi (d-5) k T}{r_+} +  \frac{(4 \pi)^2}{12} (3d - 16) T^2 \right]   \lambda_{(i)} \bigg\} 
\end{align}
where $T = f^\prime(r_+)/4\pi$ is the temperature.
Note that in each case above we have set $G = 1$.  Again, if the Lagrangian contains additional terms, then the corresponding entropy densities of these terms will be simply added to the above.  

The above result applies equally well to the four dimensional case, where the only modification is the addition of the two additional terms corresponding to the contributions from $\mathcal{S}_4^{(5)}$ and $\mathcal{S}_4^{(6)}$ yielding:

\be 
s_{d=4} = \frac{r_+^{2}}{4} \bigg[1 + \frac{32 \pi^2 T^2}{r_+^4} \sum_{i=1}^6 \lambda_{(i)} \left( k + \frac{4 \pi}{3} r_+ T \right) \bigg]
\ee

A particularly noteworthy feature of the generalized quasi-topological theories is that the entropy density of a black brane is modified from the Bekenstein-Hawking result.  In Lovelock and quasi-topological gravity, while the entropy density is modified for spherical and hyperbolic black holes, the entropy of the black brane is universal to all orders in the curvature in these theories and is given simply by one quarter the horizon area, as in Einstein gravity.  In both cubic and quartic generalized quasi-topological gravity, there are modifications to this result~\cite{Hennigar:2017ego, Bueno:2017sui}.  It would be worthwhile to explore the holographic consequences of this fact in, for example, how it applies to the ratio of shear viscosity to entropy density.

\subsection{Black hole solutions in four dimensions}

In this section we aim to study the vacuum field equations and their asymptotically flat black hole solutions in four dimensions.   It is a noteworthy feature of the generalized quasi-topological theories that higher curvature corrections occur in four dimensions while maintaining relatively simple field equations. Here we only focus on the spherically symmetric solutions that can easily be generalized to topological black holes (i.e. those with $k=0$ or $-1$). The field equations naively contain up to third derivatives of the metric function, but after  imposing
spherical symmetry and setting $N=1$  they reduce to a single equation $F'=0$, where
\begin{align}
\label{Feq}
F = \frac{r}{\kappa}(k - f) + \frac{24 \kappa^2}{5} K \bigg[ &\frac{f f' f''}{r^2} \left(f - \frac{1}{2} r f' - k \right) + \frac{f'^4}{8 r} 
\nn\\
&+ \frac{f'^3}{6 r^2} \left(f + 2k \right) + \frac{ff'^2}{r^3} (k-f)  \bigg]
\end{align}
with $\kappa=8 \pi G$, $f = f(r)$ and we have set the cosmological term to zero.
We have introduced the quantity $K$ which is defined in the following way,
\be 
K = \frac{5}{6 \kappa^3} \sum_{i=1}^{6} \lambda_{(i)}
\ee
note when $K=0$ the the field equation reduces to the Einstein gravity case.  In many of the following equations we shall keep factors of $k$ visible, since they serve as useful accounting devices, but at the end of the calculation we will set $k = 1$ to study the spherical asymptotically flat black hole.

  After integrating the field equation we get
\beqa
F=\frac{C}{\kappa}
\eeqa
where $C$ is the integration constant and the factor $1/\kappa$ gives us the valid contribution to the mass coming from the large $r$ solution, as we will see shortly.

The field equation is not solvable exactly therefore we construct a perturbative solution. We consider the asymptotic flat solution, hence as $r\rightarrow \infty$, assuming $K$ terms to be small corrections, the expansion of the metric function reads
\beqa
f(r)=k-\frac{C}{r}+\epsilon h(r)
\eeqa
where $\epsilon$ indicates the order of contribution of $h(r)$. We substitute above expansion into Eq.~\reef{Feq} and only keep the linear terms in $h(r)$. This way, we get
a second order inhomogeneous differential equation where we set $\epsilon=1$. Up to the first order in $K$, a particular solution is given by
\beqa
h_p(r)=k-\frac{C}{r}+\frac{108 \kappa^3 k   K C^3}{5 r^9}-\frac{97 \kappa^3  K C^4}{5 r^{10}}+\cO\left(\frac{K^2 C^5 }{r^{17}}\right)
\eeqa

The homogeneous equation takes the form,
\beqa
h''_h-\frac{5}{r}h'_h-\omega^2 r^6 h_h=0
\eeqa
where
\beqa
\omega^2=\frac{5 }{36\kappa^3 k C^2  |K| }
\eeqa
Here we assume $\omega^2$ is positive, which requires that $K$ is negative. This equation can be solved exactly in terms of Bessel functions, but here the relevant behaviour can be captured using an approximate solution. For large $r$, the first derivative term is negligible and the solution is approximately,
\beqa
h_h(r)\approx A \exp\left(\frac{\omega r^4}{4}\right)+B \exp\left(-\frac{\omega r^4}{4}\right)
\eeqa
For asymptotic flat spacetime we get $A=0$, so at leading order we obtain
\beqa
h(r)\approx h_p(r)+B \exp\left(-\frac{\omega r^4}{4}\right)
\label{farsol}
\eeqa
The homogeneous solution is similar to Yukawa-type terms and exponentially decaying, thus it can be neglected and we left with the particular solution as the correction. Also it is justified from the fact that the theory does not have massive modes in its spectrum.

For the asymptotically flat solutions, the ADM mass is given by \cite{Deser:2002jk}
\beqa
M = \frac{d-2}{2 \kappa } \omega_{(k)d-2} \lim_{r \to \infty} r^{d-3}(k- g_{tt})=\frac{\omega_{(k)2}C}{8 \pi G} 
\label{ADMmass}
\eeqa
where $\omega_{(k)d-2}$ is the volume of the space with the line element $d\Sigma_{(k)d-2}$; for a two-sphere this is just $\omega_{2} = 4 \pi$.

We also study the behaviour of the solution near the event horizon by expanding the metric function as
\beqa
f(r)=4\pi T(r-r_+)+\sum_{n=2}a_n (r-r_+)^n
\label{nearsol}
\eeqa
where $T=f'(r_+)/(4\pi)$ is the Hawking temperature and we use temperature instead of $f'(r_+)$. By inserting this expansion into the field equation \reef{Feq} and performing series expansion in $(r-r_+)$, at zero and first orders we find following relations
\beqa
\label{eqn:NH_expansion}
\frac{C}{\kappa}&=&\frac{1}{5 \kappa r_+^2}\left(5 k r_+^3+512 \pi ^3 k K \kappa^3 T^3+768 \pi ^4 K \kappa^3 r_+ T^4\right)\nonumber\\
0&=&\frac{1}{5 \kappa r_+^3}\left(5 k r_+^3-20 \pi  r_+^4 T+512 \pi ^3 k K \kappa^3 T^3+256 \pi ^4 K \kappa^3 r_+ T^4\right)
\eeqa
These equations determine $C$ that is related to the mass according \reef{ADMmass}, and $T$ in terms of the horizon radius, $r_+$. The second equation is quartic in $T$ but only one of its roots approaches real nonnegative value as $K\rightarrow 0$.  This is the appropriate branch to take since it has a smooth Einstein limit  and we shall use it in what  follows.

\begin{figure*}[htp]
\centering
\begin{tabular}{cc}
\includegraphics[scale=.4]{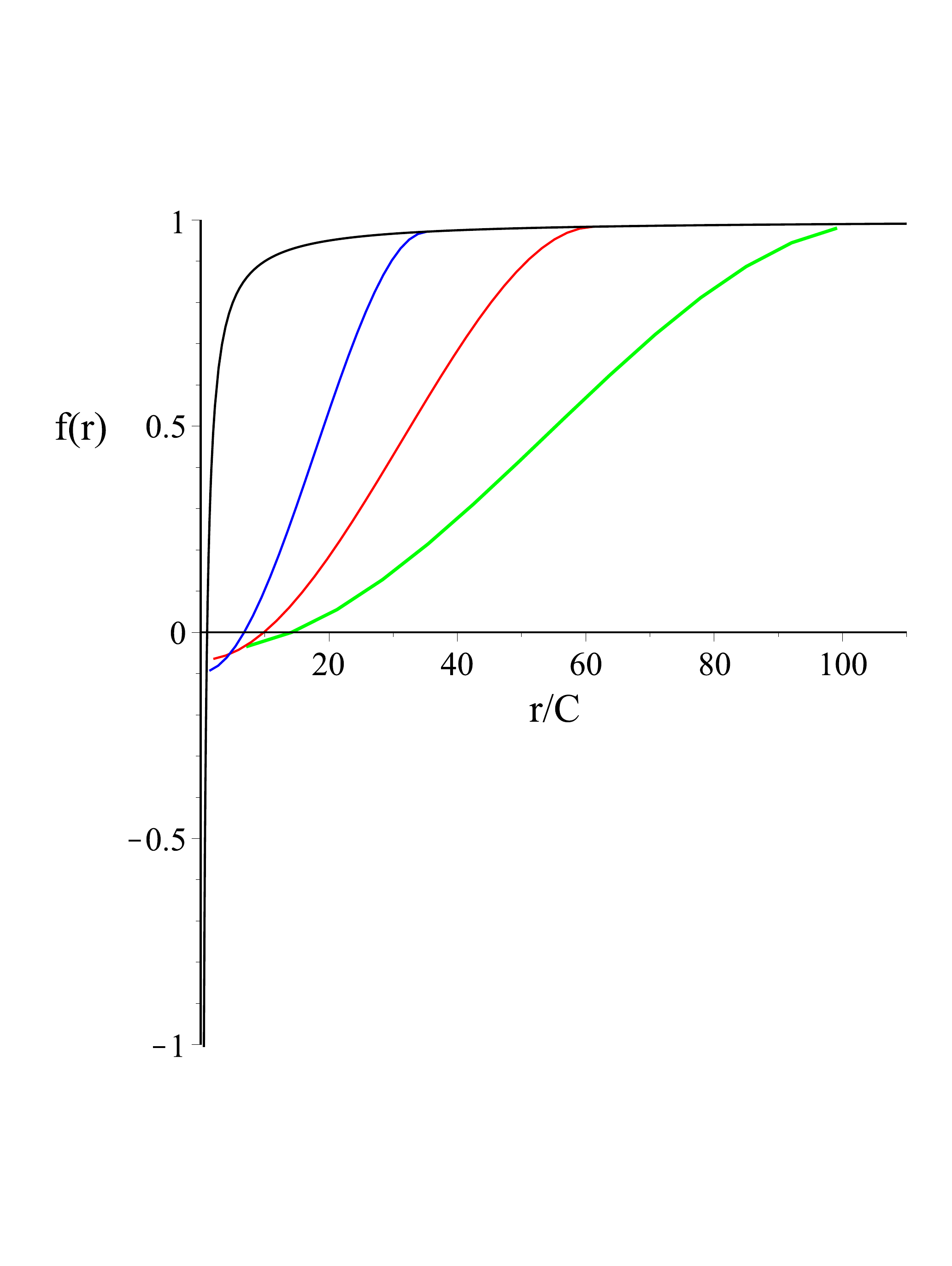}
 \\
\end{tabular}
\caption{{\bf Numerical solution} (color online).  A plot showing numeric solutions for an asymptotically flat black hole in the quartic theory.  The black curve corresponds to the usual Schwarzschild solution of Einstein gravity. The blue, red and green curves correspond to $K=-0.32, -1, -3$ respectively.}
\label{asymf}
\end{figure*}

 With near horizon and asymptotic solutions constructed, we now  join the two together by numerically solving the field equation.  The essential idea is to evaluate the near horizon expansion very near to the horizon and use this as initial data for the numerical scheme. For this purpose one should include higher orders in the expansion \reef{nearsol}. Although the higher order terms are more cumbersome, it turns out that at each order one can solve for the new parameter $a_n$ in terms of parameters in previous orders that themselves are eventually related to the single  free parameter $a_2$ at second order. We consider the value of this free parameter as~\cite{Lu:2015cqa}
\beqa
a_2=\frac{f''(r_+)}{2}=-\frac{1}{r_+^2}[1+\delta]
\eeqa
where $\delta$ states the amount of correction with respect to the Schwarzschild solution where $\delta=0$.  The value of $\delta$ must be carefully chosen for consistency with the boundary conditions (i.e. $f(r) \to 1$ as $r \to \infty$) and we use the shooting method to determine  it.  In the calculations we used terms up to order $(r-r_+)^{12}$ in the near horizon series expansion.   Good results can be obtained with fewer terms, but the construction these terms is easily automated and therefore working to a high order comes with no extra difficulty.  We have found that determining $\delta$ to ten significant digits is sufficient to integrate the solution to the point where the large $r$ expansions become accurate (see Fig~\ref{asymf}). 

The numeric results are presented in Fig. \ref{asymf}, where for various values of $K$, we exhibit $f(r)$ in terms of a dimensionless distance in four dimensions $r/C$. The graph shows that the solution approaches the expected value for asymptotically flat black holes. For larger quartic coupling we get more outward displacement of the horizon compared to the Schwarzschild black hole. In fact the presence of the higher curvature terms removes metric singularity as $r\rightarrow 0$. However, the curvature singularity remains, with the Kretschmann scalar diverging as $1/r^4$ and Ricci scalar as $1/r^2$ as can be confirmed through an expansion of $f(r)$ as $r \to 0$.  These results are similar to what was observed in the cubic case~\cite{Bueno:2016lrh, Hennigar:2017ego}.

It is worth highlighting the fact that, despite the equation of motion for $f(r)$ in this class of theories being a third order differential equation (albeit a total derivative), asymptotically the black hole solutions we have obtained are characterized only by their mass.  In the near horizon solution, the free parameter $a_2$ is equivalent to a choice of boundary conditions, and it appears that their is a unique value yielding asymptotically flat conditions. One might naively expect that since the equations of motion are third order that the black holes may possess ``higher derivative hair" (see, e.g.~\cite{Goldstein:2017rxn} for a recent discussion).  However, the above discussion shows that this is clearly not the case for this class of theories.  The black holes are characterized by a single free parameter (after fixing boundary conditions) and that is the black hole mass.  In~\cite{Bueno:2017sui} it was suggested that this may be a general feature of this class of theories.

As a consistency check of the calculations, one can verify that the first law of black hole thermodynamics holds.  Using Eq.~\eqref{eqn:NH_expansion}, and replacing $C$ with $M$ from Eq.~\eqref{ADMmass}, we find that
\be 
\delta M = T \delta S
\ee
where
\be 
S = \frac{\omega_{2} r_+^{2}}{2 \kappa} \bigg[1 + \frac{192 \pi^2 T^2}{5 r_+^4}\kappa^3 K \left( k + \frac{4 \pi}{3} r_+ T \right) \bigg]
\ee
from above.  Since each of the terms going into the first law was computed independently, the fact that this relationship holds provides an important check of our calculations.  

\subsection{Black branes}

We next consider black brane solutions of the quartic generalized quasi-topological theories.  We consider only the Einstein-Hilbert term supplemented by the quartic generalized quasi-topological terms to see more directly the effects of these terms.  We employ the following metric
\be 
ds^2 = \frac{r^2 }{\ell^2 } \left[-N(r)^2 f(r) dt^2 + \sum_{i=1}^{d-2} dx_i^2 \right] + \frac{\ell^2 dr^2}{r^2 f(r)}
\ee
in terms of which the field equations read $F' = 0$ with
\begin{align}
F =&  \left( d-2 \right) \left( r \ell \right) ^{d-3}  \big[ {r}^{2} \left(f -1  \right)   \big] + \frac{\left( d-2 \right) \lambda}{4}{\ell}^{d-3}  {r}^{d-1} \bigg[  \left(  \left( 
3d-16 \right) rf'+4\left( d-6 \right) f   
 \right) {r}^{3}  f  f' 
 \nn\\
 &-\frac{1}{4} \left( 3d-
16\right) {r}^{4}{f'}^{4}+ \frac{\left( 3d-16 \right)  \left( d+1
 \right)}{3}  {r}^{3}  f  {f'}^{3}+2d\left( d-6 \right) f^2 {r}^{2}
  {f'}^{2}+\frac{4{f}^{4}}{3} \left( d-8 \right) 
 \bigg]
\end{align}
and we have set $k=0$ and rescaled $\lambda$ with powers of $\ell$ so that it is dimensionless. Integrating we obtain $F = C$, where $C$ is an integration constant to be related to the mass.  In this case, we will set $N = 1/\sqrt{f_\infty}$ which, from a holographic perspective, is the statement that the speed of light in the dual CFT is equal to unity.

As with the spherical case presented above, we were unable to solve the field equations exactly.  We therefore employ perturbative methods here.  Considering first the asymptotic solution, we write
\be 
f(r) = f_{\infty} - \frac{\ell^2 C}{(r\ell)^{d-1} } + \epsilon h(r) \, .
\ee
The quantity $f_{\infty}$ is the asymptotic value of the metric function which is a solution of the quartic polynomial equation,
\be 
1 - f_\infty + \frac{(d-8)}{3} \lambda f_\infty^4 = 0 \, .
\ee
Defined this way, the black branes asymptote to an AdS space with curvature radius $\tilde{\ell} = \ell/\sqrt{f_\infty}$.

We plug this ansatz for $f(r)$ into the field equations, set all terms proportional to $\epsilon^n$ equal to zero for $n > 1$ and then finally set $\epsilon = 1$ to obtain an inhomogenous second order differential equation for the correction $h(r)$.  The general form of this expression is complicated, and so we do not present it here.  However, it is easy to show that a particular solution takes the form,
\begin{align}
h_p(r) &= \frac{4(d-8) f_\infty^3 \lambda}{4(d-8) f_\infty^3 \lambda - 3} \frac{\ell^2 C}{(r \ell)^{d-1} }
\nn\\
& - \frac{3 (d^4 - 8 d^3 + 13 d^2 - 10d + 32)\left(4(d-8) \lambda f_\infty^3 + 3\right)  \lambda f_\infty^2}{2 \left(4(d-8) \lambda f_\infty^3 - 3\right)^2}  \frac{\ell^4 C^2}{(r\ell)^{2d-2}} + {\cal O} \left( \frac{\lambda C^3}{r^{3d-3}}\right)
\end{align}
We note that, provided $f_\infty \neq 0$, there are  corrections of the same order as the mass of the black brane from the higher curvature terms.

The form of the homogenous equation, at large $r$, must be considered in two separate cases.  First, we consider the  $d \neq 6$ case.  There we have,
\be 
h'' -  {\frac { \left( 3\,d-16 \right)  \left( d-1 \right) ^{2} C}{4 \left( d-6 \right) f_{\infty} r^d \ell^{d-3} }} h' - \omega^2_d {r}^{d-3} h = 0 \, , \quad \text{(for $d \neq 6$)}
\ee
with
\be 
\omega^2_d = {\frac {3 - 4\left( d-8 \right)\lambda f_\infty^{3} }{3\lambda\, \left( d-1 \right) \left( d-6 \right) C f_\infty^
{2}  }} \ell^{d-3} \, \, .
\ee
To keep with the AdS boundary conditions and avoid oscillating solutions we must have $\omega^2 > 0$ which constrains the coupling to satisfy,
\be 
\frac{3 - 4(d-8) \lambda f_\infty^3}{(d-6) \lambda } > 0 \, .
\ee
Physically reasonable solutions in must satisfy this inequality.  Provided this condition is met, an approximate solution to the homogenous equation is given by,
\be 
h(r) \approx A \exp \left( \frac{2 \omega_d r^{(d-1)/2}}{d-1} \right) + B \exp\left(-\frac{2 \omega_d r^{(d-1)/2}}{d-1} \right) \, .
\ee 
We of course set $A = 0$ to maintain consistency with our boundary conditions and also note that the second term is hugely suppressed and so can be discarded.  

When the calculation is done in six dimensions, the homogeneous equation reads,
\be 
h'' - \frac{5}{r} h' - \omega^2_{6} r^8 h = 0
\ee
where
\be 
\omega_{6}^2 = \frac{2(3 + 8 \lambda f_\infty^3) }{75 \lambda C^2 f_\infty} \ell^6
\ee
and we must have
\be 
\frac{3 + 8 \lambda f_\infty^3}{\lambda} > 0 \,  \quad \text{for $d = 6$}.
\ee 
The approximate solution then reads,
\be 
h(r) \approx A \exp \left( \frac{ \omega_6 r^{5}}{5} \right) + B \exp\left(-\frac{\omega_6 r^{5}}{5} \right) \, .
\ee
We again set $A=0$ in this case for consistency with the boundary conditions, and discard the second term since it is enormously suppressed relative to the particular solution.

The above considerations reveal restrictions which must be enforced on the couplings that are not clear from the particular solution alone.  Further, we see that we are permitted to drop the exponential terms from our considerations.

Next we turn to the near horizon solution where we expand the metric function as,
\be 
f(r) = \frac{4 \pi T \sqrt{f_\infty} \ell^2}{r_+^2}  (r- r_+) + \sum_{i=2} a_n (r - r_+)^n
\ee
assuming that the metric function vanishes linearly as $r \to r_+$  for non-extremal black holes.  Substituting this ansatz into the field equations, one arrives at recurrence relations determining the coefficients in the series expansion.  The first two of these are
\begin{align}
C &=  \ell^{d-3} r_+^{d-1} - 16 (3d - 16) \pi^4 \lambda r_+^{d-5} \ell^{d+5}  f_{\infty}^2 T^4 \, ,
\nn\\
0 &= (d-1) r_+^{4} - 4 \pi \sqrt{f_\infty} T r_+^{3} \ell^{2}  + \frac{16 \lambda }{3} (d-5)(3d-16)f_\infty^2 \ell^{8} \pi^4 T^4
\label{eqn:black_brane_NH}
\end{align} 
and allow us  to determine the temperature and integration constant $C$ in terms of the horizon radius.  The higher order terms in the series solution can be easily computed, but they rapidly become increasingly complex and are not actually needed to study the thermodynamics.  As has been observed in several instances now~\cite{Hennigar:2016gkm, Bueno:2016lrh, Hennigar:2017ego, Bueno:2017sui} the thermodynamics of black objects in the generalized quasi-topological theories can be studied exactly despite the lack of an exact solution to the full field equations.  

An interesting feature of the above result is that in five dimensions, the quartic generalized quasi-topological terms do not modify the temperature from the Einstein gravity result. However, in all other dimensions, the temperature gets corrected by these terms.  

Recasting the entropy density presented earlier into the notation used for the black branes, we arrive at
\be 
s = \frac{1}{4} \left(\frac{r_+}{\ell}\right)^{d-2} \bigg[1 - \frac{16\lambda}{3} \frac{(d-2)(3d-16)\pi^3 \ell^6 f_\infty^{3/2} T^3}{r_+^3} \bigg]  
\ee
which is not simply given by the Bekenstein-Hawking area law, but rather contains corrections due to the generalized quasi-topological contributions.  This is notably different from what is observed in both Lovelock and quasi-topological gravity, where the area law remains unaffected for black branes, and may have interesting holographic consequences.  For $d\leq 5$ the entropy is larger than that
in  Einstein gravity ($\lambda=0$), whereas for $d \geq 6$ it is smaller.

It is easy to verify that the entropy density above satisfies the first law,
\be 
d\varepsilon = T ds
\ee
with the energy density given by,
\begin{align} 
\varepsilon &= \frac{(d-2)}{16 \pi \sqrt{f_\infty} \ell^{2d-3}} C \, ,
\nn\\
&= \frac{(d-2)}{16 \pi \sqrt{f_\infty} \ell^{2d-3}} \big[ \ell^{d-3} r_+^{d-1} - 16 (3d - 16) \pi^4 \lambda r_+^{d-5} \ell^{d+5}  f_{\infty}^2 T^4 \big]
\end{align}
The factors of $r_+$ appearing in the entropy and energy densities can be eliminated by solving the second equation of~\eqref{eqn:black_brane_NH}.  This is made easier by writing,
\be 
r_+ = \gamma_\lambda \frac{4 \pi \ell^2 \sqrt{f_\infty}}{d-1} T
\ee
where $\gamma_\lambda$ solves the equation,
\be\label{eqn:gamma_eq} 
\gamma_\lambda^4 - \gamma_\lambda^3 + \frac{\lambda}{48}(d-1)^3(d-5)(3d-16) = 0 \, ,
\ee
which is obtained by substituting $r_+$ for  the above definition in the second equation of~\eqref{eqn:black_brane_NH}. Here we have included the subscript $\lambda$ to illustrate that this quantity depends directly on the coupling, $\lambda$.
The entropy and energy densities can then be recast as,
\begin{align}
s &= \frac{12 \gamma_\lambda^3 - \lambda (d-1)^3(d-2)(3d-16)}{48 \gamma_\lambda^3} \left(\gamma_\lambda \frac{4 \pi \ell  \sqrt{f_\infty} T}{d-1} \right)^{d-2} \, ,
\nn\\
\varepsilon &= \frac{(d-2) \ell^{d-4}}{256\pi  \gamma_\lambda^4}\left[16 \gamma_\lambda^4 - (d-1)^4(3d - 16)\lambda  \right] \left(\gamma_\lambda \frac{4 \pi \ell^2 \sqrt{f_\infty}}{d-1} T \right)^{d-1} \,.
\end{align}
By studying the polynomial \eqref{eqn:gamma_eq}, we can conclude that there will be real, positive solutions for $\gamma_\lambda$ provided the coupling satisfies,
\be 
\lambda \le \frac{81}{16(d-1)^3(d-5)(3d-16)}
\ee
with equality corresponding to a positive, real double root.  Of course, the above constraint does not apply in $d = 5$, but in this case, $\lambda$ does not contribute to the polynomial, and the only valid solution is $\gamma_\lambda = 1$, which holds for any value of the coupling.

From the entropy and energy densities, one can construct the free energy density,  which is given by,
\be 
\mathcal{F} =  \varepsilon - T s = - \frac{12 \gamma_\lambda^3 - \lambda (d-1)^3(d-2)(3d-16)}{192 \pi \ell \sqrt{f_\infty} \gamma_\lambda^4} \left(\gamma_\lambda \frac{4 \pi \ell  \sqrt{f_\infty} T}{d-1} \right)^{d-1} \, .
\ee
The entropy and energy densities can be shown [using Eq.~\eqref{eqn:gamma_eq}] to satisfy the relation,
\be 
\varepsilon = \frac{d-2}{d-1} Ts
\ee
as expected for a CFT living in $d-1$ dimensions.

An interesting aspect of the above results is that the entropy and energy densities are modified from the Einstein gravity result. Similar results were noted in~\cite{Bueno:2017sui} for five dimensional black branes in cubic generalized quasi-topological gravity. In Lovelock and quasi-topological gravity, this is not the case: the expressions are identical, apart from the appearance of the term $f_\infty$ characterizing the curvature of the AdS space~\cite{Myers:2010ru}.  In a sense, the properties of black branes in these latter theories are `universal'.  

We expect that differences appearing in the generalized quasi-topological case will lead to further interesting results when a full holographic study of these theories is performed.

\section{Conclusions}

We have constructed a complete set of theories of gravity that are quartic in curvature and whose field equations each reduce to a total
derivative of a functional of one metric function under the restriction to spherical symmetry.  In four dimensions, the non-trivial contributions arise from six generalized quasi-topological theories given by~\eqref{eqn:4d_GQT} and~\eqref{eqn:picking_GQT} which have  equations of motion that are a total derivative of a polynomial of $f(r)$ and its first two derivatives. In five or higher dimensions, these theories break up into the following classes:

\begin{enumerate}
\item  4-th order Lovelock gravity, whose Lagrangian is given by the eight dimensional Euler density $\mathcal{X}_8$.  The field equations of this theory are
always second-order, and under restriction to spherical symmetry becomes one equation that is a total derivative of a polynomial in a single metric function $f$.
\item  Six quasi-topological theories, whose Lagrangians are given by \eqref{qZ-1} and \eqref{qZ-2}. The theory described by 
\eqref{qZ-1} was found previously~\cite{Dehghani:2011vu}; the remaining five in \eqref{qZ-2} are new.  For all six, under restriction to spherical symmetry 
the field equations become a single equation that is a total derivative of a polynomial in a single metric function $f$. In the context of spherical symmetry, the field equations of all six theories coincide, since the Lagrangians are equivalent up to terms which vanish for \textit{spherically symmetric} metrics. Relaxing the constraint of spherical symmetry, the field equations of the theories will no longer coincide (since they are distinct Lagrangians) and are fourth order differential equations.
 \item A quartet  of generalized quasi-topological theories, whose Lagrangians are given by \eqref{eqn:picking_GQT}.  For all four, under restriction to spherical symmetry  the field equations become a single equation that is a total derivative of a polynomial in a single metric function $f$  and its first and second derivatives. 
\item Six theories,  whose Lagrangians are given by \eqref{C-theories}, and for which the field equations vanish if the function
$N$ is constant. For situations where the stress-energy  $T_{t}^t \neq T_{r}^r$ there will be two non-trivial field equations 
that determine $N$ and $f$.
\end{enumerate}

We have also studied several aspects of these theories.  We have considered their linearized spectrum, finding that it is only a massless, transverse graviton that is propagated on a constant curvature background.  We have presented the field equations of the theories, valid for a static spherically symmetric metric in any dimension, and have determined the expression for black hole entropy  in arbitrary dimensions.  This latter result is particularly interesting since, for black branes, there are corrections to the Bekenstein-Hawking area law, something which does not occur in either Lovelock or quasi-topological gravity, and may have interesting holographic consequences.  Studying four dimensional asymptotically flat black hole solutions, we have found that the solutions are characterized simply by their mass and are free from any higher derivative hair.  In this case, the thermodynamics can be studied exactly (despite having only perturbative and numeric solutions) and we find the first law holds.  Black brane solutions of the theory were studied in arbitrary dimensions.  These solutions satisfy the expected thermodynamic relations for a CFT (without chemical potential) living in one dimension less.  Further, unlike the case in Lovelock or quasi-topological gravity, the generalized quasi-topological terms modify the thermodynamic properties of the black branes.  This result may have interesting consequences in holographic studies.

This class of theories (which has now been constructed to cubic~\cite{Bueno:2016xff, Hennigar:2017ego} and quartic order) provides interesting generalizations of Einstein gravity that  are non-trivial in four (and higher) dimensions.  This contrasts with previous constructions of Lovelock and quasi-topological gravity, which vanish on four dimensional (spherically symmetric) metrics. The generalized quasi-topological terms can be thought of as the theories which have many of the interesting properties observed for Einsteinian cubic gravity~\cite{Bueno:2016xff} in four dimensions~\cite{Hennigar:2016gkm, Bueno:2016lrh}, but in higher dimensions and/or to higher orders in the curvature.   These theories necessarily~\cite{Bueno:2017sui} propagate only a massless, transverse graviton on a maximally symmetric vacuum.  Furthermore, they admit black hole solutions which are characterized only by their mass.  The thermodynamics of the black holes can be studied exactly despite the lack of an exact, analytic solution to the field equations.

There remain many problems deserving  further study.      It would be worthwhile to further study the properties of the four and higher dimensional black holes in these theories, and work is currently in progress on this~\cite{HennigarPrep}.  It would be useful to know if the Birkhoff theorem holds for the generalzied quasi-topolgoical terms, similar to Lovelock and quasitopological gravities~\cite{Oliva:2011xu, Oliva:2012zs, Ray:2015ava}. More generally, the generalized quasi-topological theories seem well suited for holographic study and may serve as useful toy models in this context. A holographic study would also shed further light on stability and the permissible values of coupling constants, and may reveal novel features in the case of black brane solutions of the theory. An ambitious undertaking would be to elucidate the general structure of the Lagrangians in this class of theories.  This has been long known in the case of Lovelock gravity~\cite{Lovelock:1971yv}, but remains an open problem in the (generalized) quasi-topological cases.

\section*{Acknowledgements}
The work of Jamil Ahmed is supported by the fellowship of Pakistan Higher Education Commission (HEC) under its international research initiative program and also a research grant from the HEC under its Project No. 20-2087 is gratefully acknowledged.  This work was supported in part by the Natural Sciences and Engineering Research Council of Canada.

\appendix

\section{The constraints in general dimensions}
\label{app:constraints}
The following constraints on $c_{12}$, $c_{17}$, $c_{19}$, $c_{20}$, $c_{21}$, $c_{22}$, $c_{23}$, $c_{24}$ and $c_{25}$ ensure  that condition~\eqref{eqn:feq_condition} is met for the quartic action in dimensions larger than four.
\begin{align} 
c_{12} &= 
-  \frac{ (19 - 40 d + 38 d^2 - 15 d^3 + 2 d^4)}{3 
(3d - 2) (-22 + 26 d - 9 d^2 + d^3)}c_1 -  \frac{ (2 - 69 d + 83 
d^2 - 32 d^3 + 4 d^4)}{3 (3d - 2) (-22 + 26 d - 9 d^2 + d^3)} c_2
\nn\\
&-  
\frac{ (13 - 2 d - 4 d^2 + d^3)}{3 (3d - 2) (-22 + 26 d - 9 d^2 
+ d^3)} c_3 -  \frac{4  (d - 2) (-2 + 22 d - 13 d^2 + 2 d^3)}{3 (3d -2) (-22 + 26 d - 9 d^2 + d^3)} c_4 
\nn\\
& -  \frac{8  (d - 2) (-2 + 22 d - 
13 d^2 + 2 d^3)}{3 (3d - 2) (-22 + 26 d - 9 d^2 + d^3)} c_5  -  \frac{8 (d - 3) (-1 + 6 d - 5 d^2 + d^3)}{3 (3d - 1) (-22 + 26 d - 9 d^2 + d^3)}c_6 
\nn\\
&- \frac{ (d -4 ) (d -3) (d -1)^2}{2 (3d -2 ) (-22 + 26 d - 9 d^2 + d^3)} c_{10}  -  \frac{8 (d - 3) (d - 2)^2 (2d - 1)}{3 (3 d - 2) (-22 + 26 d - 9 d^2 + d^3)} c_7 
\nn\\
&-  \frac{(d - 2) (1 - 7 d + 2 
d^2)}{4 (3d - 2) (-22 + 26 d - 9 d^2 + d^3)}c_8 -  
\frac{ (5 - 28 d + 27 d^2 - 9 d^3 + d^4)}{(3d - 2) (-22 + 26 d - 
9 d^2 + d^3)}c_9 
\nn\\
&  
-  \frac{(16 - 15 
d + 3 d^2)}{4 (-22 + 26 d - 9 d^2 + d^3)} c_{11}
\end{align}

\begin{align}
c_{17} &=-  \frac{(-200 + 430 d + 566 d^2 - 2677 d^3 + 3194 d^4 - 1807 d^5 + 524 d^6 - 74 d^7 + 4 d^8)}{2 (d-2)^2 (d-1) (2d-1) (3d - 2) (-22 + 26 d - 9 d^2 + d^3)}c_1 
\nn\\
&-  \frac{ (272 - 1572 d + 4104 d^2 - 5617 d^3 + 4420 d^4 - 2042 d^5 + 536 d^6 - 73 d^7 + 4 d^8)}{(d-2)^2 (d-1) (2d-1) (3d - 2) (-22 + 26 d - 9 d^2 + d^3)}c_2
\nn\\
&+ \frac{(336 - 1418 d + 2520 d^2 - 2107 d^3 + 885 d^4 - 168 d^5 + 7 d^6 + d^7)}{2 (d-2)^2 (d-1) (2d-1) (3d - 2) (-22 + 26 d - 9 d^2 + d^3)}c_3
\nn\\
&-  \frac{4 (148 - 790 d + 1986 d^2 - 2683 d^3 + 2126 d^4 - 991 d^5 + 262 d^6 - 36 d^7 + 2 d^8)}{(d-2)^2 (d-1) (2d-1) (3d - 2) (-22 + 26 d - 9 d^2 + d^3)}c_4
\nn\\
&-  \frac{8 (148 - 790 d + 1986 d^2 - 2683 d^3 + 2126 d^4 - 991 d^5 + 262 d^6 - 36 d^7 + 2 d^8)}{(d-2)^2 (d-1) (2d-1) (3d - 2) (-22 + 26 d - 9 d^2 + d^3)}c_5
\nn\\
&-  \frac{2 (136 - 988 d + 3086 d^2 - 4784 d^3 + 4079 d^4 - 1975 d^5 + 531 d^6 - 73 d^7 + 4 d^8)}{(d-2)^2 (d-1) (2d-1) (3d - 2) (-22 + 26 d - 9 d^2 + d^3)}c_6
\nn\\
&-  \frac{8  (20 + 58 d - 134 d^2 + 74 d^3 - 15 d^4 + d^5)}{(d-2) (3d - 2) (-22 + 26 d - 9 d^2 + d^3)}c_7
\nn\\
&+ \frac{(24 + 36 d - 296 d^2 + 683 d^3 - 698 d^4 + 368 d^5 - 94 d^6 + 9 d^7)}{4 (d-2)^2 (d-1) (2d-1) (3d - 2) (-22 + 26 d - 9 d^2 + d^3)}c_8
\nn\\
&-  \frac{3 (40 - 546 d + 2232 d^2 - 3935 d^3 + 3633 d^4 - 1870 d^5 + 538 d^6 - 81 d^7 + 5 d^8)}{2 (d-2)^2 (d-1) (2d-1) (3d - 2) (-22 + 26 d - 9 d^2 + d^3)}c_9
\nn\\
&-  \frac{(-240 + 184 d + 2538 d^2 - 7062 d^3 + 7893 d^4 - 4550 d^5 + 1418 d^6 - 228 d^7 + 15 d^8)}{4 (d-2)^2 (d-1) (2d-1) (3d - 2) (-22 + 26 d - 9 d^2 + d^3)} c_{10}
\nn\\
&+ \frac{3 (d - 4) (d-3) (d-1)^2 (-1 + 3 d)}{2 (d-2)^2 (2d-1) (-22 + 26 d - 9 d^2 + d^3)}c_{11}  -  \frac{ (-4 - 13 d + 39 d^2 - 24 d^3 + 4 d^4)}{4 (d-2)^2 (d-1) (2d-1)} c_{13}     
\nn\\
&- \frac{ (d-3) d^2}{4 (d-2)^2 (d-1)}c_{14} -  \frac{ (-2 - 8 d + 23 d^2 - 13 d^3 + 2 d^4)}{2 (d-2)^2 (d-1) (2d-1)} c_{15}  
\nn\\
&-  \frac{ (2 + 7 d - 9 d^2 + 2 d^3)}{2 (d-2)^2 (2d-1)}  c_{16} -  \frac{(d - 4) (d-1) (-1 + 3 d)}{2 (d-2)^2 (2d-1)} c_{18}  
\end{align}

\begin{align}
c_{19} &= \frac{(-344 + 2086 d - 4878 d^2 + 5109 d^3 - 2618 d^4 + 590 d^5 - 10 d^6 - 17 d^7 + 2 d^8)}{4 (d-2)^2 (d-1) (2d-1) (3d - 2) (-22 + 26 d - 9 d^2 + d^3)}c_1
\nn\\
&+ \frac{(-1000 + 4684 d - 7926 d^2 + 6691 d^3 - 2963 d^4 + 595 d^5 -  d^6 - 18 d^7 + 2 d^8)}{2 (d-2)^2 (d-1) (2d-1) (3d - 2) (-22 + 26 d - 9 d^2 + d^3)}  c_2
\nn\\
&+ \frac{(552 - 2214 d + 2702 d^2 - 1489 d^3 + 358 d^4 - 14 d^5 - 8 d^6 + d^7)}{4 (d-2)^2 (d-1) (2d-1) (3d - 2) (-22 + 26 d - 9 d^2 + d^3)}  c_3
\nn\\
&+ \frac{(-1096 + 5100 d - 8504 d^2 + 7072 d^3 - 3046 d^4 + 579 d^5 + 8 d^6 - 19 d^7 + 2 d^8)}{(d-2)^2 (d-1) (2d-1) (3d - 2) (-22 + 26 d - 9 d^2 + d^3)} c_4
\nn\\
&+ \frac{2 (-1096 + 5100 d - 8504 d^2 + 7072 d^3 - 3046 d^4 + 579 d^5 + 8 d^6 - 19 d^7 + 2 d^8)}{(d-2)^2 (d-1) (2d-1) (3d - 2) (-22 + 26 d - 9 d^2 + d^3)} c_5
\nn\\
&+ \frac{2 (-448 + 2150 d - 3790 d^2 + 3299 d^3 - 1488 d^4 + 302 d^5 -  d^6 - 9 d^7 + d^8)}{(d-2)^2 (d-1) (2d-1) (3d - 2) (-22 + 26 d - 9 d^2 + d^3)} c_6
\nn\\
&+ \frac{2 (368 - 580 d + 262 d^2 - 30 d^3 - 5 d^4 + d^5)}{(d-2) (3d - 2) (-22 + 26 d - 9 d^2 + d^3)}  c_7
\nn\\
&+ \frac{(360 - 1388 d + 1568 d^2 - 593 d^3 - 124 d^4 + 146 d^5 - 36 d^6 + 3 d^7)}{8 (d-2)^2 (d-1) (2d-1) (3d - 2) (-22 + 26 d - 9 d^2 + d^3)} c_8
\nn\\
&+ \frac{3 (-776 + 3662 d - 6388 d^2 + 5545 d^3 - 2543 d^4 + 576 d^5 - 38 d^6 - 7 d^7 + d^8)}{4 (d-2)^2 (d-1) (2d-1) (3d - 2) (-22 + 26 d - 9 d^2 + d^3)} c_9
\nn\\
&+ \frac{(-1152 + 6104 d - 12474 d^2 + 12330 d^3 - 6321 d^4 + 1592 d^5 - 128 d^6 - 18 d^7 + 3 d^8)}{8 (d-2)^2 (d-1) (2d-1) (3d - 2) (-22 + 26 d - 9 d^2 + d^3)} c_{10}
\nn\\
&+ \frac{3 (d-3) (16 - 49 d + 41 d^2 - 11 d^3 + d^4)}{4 (d-2)^2 (2d-1) (-22 + 26 d - 9 d^2 + d^3)} c_{11}  + \frac{(-11 + 25 d - 14 d^2 + 2 d^3)}{4 (d-2)^2 (d-1) (2d-1)} c_{13}     
\nn\\
&+ \frac{ (d-3) d}{4 (d-2)^2 (d-1)} c_{14} + \frac{ (-7 + 16 d - 8 d^2 + d^3)}{2 (d-2)^2 (d-1) (2d-1)} c_{15}+ \frac{ (d - 4) (d-1)}{2 (d-2)^2 (2d-1)} c_{16}  
\nn\\
&-  \frac{(-8 + 17 d - 6 d^2 + d^3)}{4 (d-2)^2 (2d-1)} c_{18}
\end{align}

\begin{align}
c_{20} &=  -  \frac{8 (66 - 106 d - 27 d^2 + 99 d^3 - 52 d^4 + 8 d^5)}{(d-2) (2d-1) (3d - 2) (-22 + 26 d - 9 d^2 + d^3)} c_1
\nn\\
&-  \frac{8 (-220 + 638 d - 716 d^2 + 427 d^3 - 133 d^4 + 16 d^5)}{(d-2) (2d-1) (3d - 2) (-22 + 26 d - 9 d^2 + d^3)} c_2
\nn\\
&-  \frac{4 (264 - 634 d + 494 d^2 - 175 d^3 + 23 d^4)}{(d-2) (2d-1) (3d - 2) (-22 + 26 d - 9 d^2 + d^3)} c_3
\nn\\
&-  \frac{32 (-132 + 362 d - 384 d^2 + 227 d^3 - 69 d^4 + 8 d^5)}{(d-2) (2d-1) (3d - 2) (-22 + 26 d - 9 d^2 + d^3)} c_4 
\nn\\
&-  \frac{64 (-132 + 362 d - 384 d^2 + 227 d^3 - 69 d^4 + 8 d^5)}{(d-2) (2d-1) (3d - 2) (-22 + 26 d - 9 d^2 + d^3)} c_5
\nn\\
&-  \frac{16 (-132 + 422 d - 548 d^2 + 373 d^3 - 127 d^4 + 16 d^5)}{(d-2) (2d-1) (3d - 2) (-22 + 26 d - 9 d^2 + d^3)} c_6
\nn\\
&-  \frac{256 (d-3) (d-1) d}{(3d - 2) (-22 + 26 d - 9 d^2 + d^3)} c_7 -  \frac{4 (44 - 80 d + 26 d^2 + 13 d^3 - 14 d^4 + 3 d^5)}{(d-2) (2d-1) (3d - 2) (-22 + 26 d - 9 d^2 + d^3)} c_8
\nn\\
&-  \frac{12 (-44 + 116 d - 148 d^2 + 116 d^3 - 41 d^4 + 5 d^5)}{(d-2) (2d-1) (3d - 2) (-22 + 26 d - 9 d^2 + d^3)} c_9  
\nn\\
&-  \frac{2  (176 - 416 d + 242 d^2 + 22 d^3 - 53 d^4 + 9 d^5)}{(d-2) (2d-1) (3d - 2) (-22 + 26 d - 9 d^2 + d^3)}   c_{10}
\nn\\
& -  \frac{24 (d-3) (d-1)^2 d}{(d-2) (2d-1) (-22 + 26 d - 9 d^2 + d^3)} c_{11} + \frac{2  (-5 + 2 d)}{(d-2) (2d-1)} c_{13}
\nn\\
&+  \frac{4 }{d-2} c_{14} -  \frac{8 }{(d-2) (2d-1)} c_{15} + \frac{4  (-3 + 2 d)}{(d-2) (2d-1)}  c_{16} + \frac{8 (d-1) d}{(d-2) (2d-1)} c_{18}   
\end{align}

\begin{align}
c_{21} &= \frac{4 (-96 + 509 d - 1068 d^2 + 1031 d^3 - 516 d^4 + 141 d^5 - 23 d^6 + 2 d^7)}{(d-2) (d-1) (2d-1) (3d - 2) (-22 + 26 d - 9 d^2 + d^3)} c_1 
\nn\\
&+ \frac{4 (-332 + 1560 d - 2593 d^2 + 2194 d^3 - 1041 d^4 + 288 d^5 - 48 d^6 + 4 d^7)}{(d-2) (d-1) (2d-1) (3d - 2) (-22 + 26 d - 9 d^2 + d^3)} c_2
\nn\\
&+ \frac{4 (64 - 253 d + 246 d^2 - 101 d^3 + 20 d^4 - 5 d^5 + d^6)}{(d-2) (d-1) (2d-1) (3d - 2) (-22 + 26 d - 9 d^2 + d^3)} c_3
\nn\\
&+ \frac{16 (-180 + 850 d - 1408 d^2 + 1191 d^3 - 559 d^4 + 153 d^5 - 25 d^6 + 2 d^7)}{(d-2) (d-1) (2d-1) (3d - 2) (-22 + 26 d - 9 d^2 + d^3)}  c_4
\nn\\
&+ \frac{32 (-180 + 850 d - 1408 d^2 + 1191 d^3 - 559 d^4 + 153 d^5 - 25 d^6 + 2 d^7)}{(d-2) (d-1) (2d-1) (3d - 2) (-22 + 26 d - 9 d^2 + d^3)} c_5
\nn\\
&+ \frac{32 (-80 + 381 d - 657 d^2 + 566 d^3 - 268 d^4 + 73 d^5 - 12 d^6 + d^7)}{(d-2) (d-1) (2d-1) (3d - 2) (-22 + 26 d - 9 d^2 + d^3)}  c_6
\nn\\
&+ \frac{32 (d-3) (-24 + 26 d - 5 d^2 + d^3)}{(3d - 2) (-22 + 26 d - 9 d^2 + d^3)} c_7
\nn\\
&+ \frac{(116 - 464 d + 557 d^2 - 326 d^3 + 115 d^4 - 36 d^5 + 6 d^6)}{(d-2) (d-1) (2d-1) (3d - 2) (-22 + 26 d - 9 d^2 + d^3)} c_8
\nn\\
&+ \frac{12 (-148 + 689 d - 1170 d^2 + 989 d^3 - 452 d^4 + 115 d^5 - 16 d^6 + d^7)}{(d-2) (d-1) (2d-1) (3d - 2) (-22 + 26 d - 9 d^2 + d^3)}  c_9
\nn\\
&+ \frac{2 (-256 + 1284 d - 2475 d^2 + 2301 d^3 - 1132 d^4 + 304 d^5 - 45 d^6 + 3 d^7)}{(d-2) (d-1) (2d-1) (3d - 2) (-22 + 26 d - 9 d^2 + d^3)} c_{10}
\nn\\
&+ \frac{3  (-28 + 78 d - 51 d^2 + 3 d^3 + 2 d^4)}{(d-2) (2d-1) (-22 + 26 d - 9 d^2 + d^3)}  c_{11} -  \frac{2  (5 - 10 d + 4 d^2)}{(d-2) (d-1) (2d-1)}     c_{13} 
\nn\\
&- \frac{2  d}{(d-2) (d-1)} c_{14}-  \frac{4  (3 - 6 d + 2 d^2)}{(d-2) (d-1) (2d-1)} c_{15} -  \frac{8  (d-1)}{(d-2) (2d-1)} c_{16}
\nn\\
&-  \frac{4  (d-1 + d^2)}{(d-2) (2d-1)} c_{18}
\end{align}

\begin{align}
c_{22} &= \frac{ (-9504 + 28040 d - 26710 d^2 + 7806 d^3 + 2763 d^4 - 2722 d^5 + 1012 d^6 - 222 d^7 + 15 d^8 + 2 d^9)}{12 (d-2) (d-1) (2d-1) (3d - 2) (-22 + 26 d - 9 d^2 + d^3)} c_1
\nn\\
&+ \frac{(11616 - 38016 d + 46080 d^2 - 27850 d^3 + 10107 d^4 - 3153 d^5 + 1051 d^6 - 235 d^7 + 14 d^8 + 2 d^9)}{6 (d-2) (d-1) (2d-1) (3d - 2) (-22 + 26 d - 9 d^2 + d^3)} c_2
\nn\\
&+ \frac{ (-15840 + 49168 d - 53582 d^2 + 25858 d^3 - 5241 d^4 + 356 d^5 - 62 d^6 + 14 d^7 + d^8)}{12 (d-2) (d-1) (2d-1) (3d - 2) (-22 + 26 d - 9 d^2 + d^3)}c_3
\nn\\
&+ \frac{(14784 - 47304 d + 55236 d^2 - 31612 d^3 + 10792 d^4 - 3404 d^5 + 1165 d^6 - 248 d^7 + 13 d^8 + 2 d^9)}{3 (d-2) (d-1) (2d-1) (3d - 2) (-22 + 26 d - 9 d^2 + d^3)} c_4
\nn\\
&+ \frac{2 (14784 - 47304 d + 55236 d^2 - 31612 d^3 + 10792 d^4 - 3404 d^5 + 1165 d^6 - 248 d^7 + 13 d^8 + 2 d^9)}{3 (d-2) (d-1) (2d-1) (3d - 2) (-22 + 26 d - 9 d^2 + d^3)} c_5
\nn\\
&+ \frac{2 (3168 - 10564 d + 13436 d^2 - 9026 d^3 + 4002 d^4 - 1539 d^5 + 537 d^6 - 118 d^7 + 7 d^8 + d^9)}{3 (d-2) (d-1) (2d-1) (3d - 2) (-22 + 26 d - 9 d^2 + d^3)}  c_6
\nn\\
&+ \frac{2 (d-2) d (-264 + 102 d - 48 d^2 + 13 d^3 + d^4)}{3 (3d - 2) (-22 + 26 d - 9 d^2 + d^3)} c_7
\nn\\
&+ \frac{(-1760 + 4280 d - 2348 d^2 - 960 d^3 + 1127 d^4 - 150 d^5 - 76 d^6 + 14 d^7 + d^8)}{8 (d-2) (d-1) (2d-1) (3d - 2) (-22 + 26 d - 9 d^2 + d^3)} c_8
\nn\\
&+ \frac{(1408 - 1224 d - 3726 d^2 + 4544 d^3 - 245 d^4 - 1603 d^5 + 784 d^6 - 134 d^7 + 3 d^8 + d^9)}{4 (d-2) (d-1) (2d-1) (3d - 2) (-22 + 26 d - 9 d^2 + d^3)} c_9
\nn\\
&+ \frac{ (-10560 + 34480 d - 39088 d^2 + 17370 d^3 + 582 d^4 - 3803 d^5 + 1726 d^6 - 338 d^7 + 12 d^8 + 3 d^9)}{24 (d-2) (d-1) (2d-1) (3d - 2) (-22 + 26 d - 9 d^2 + d^3)} c_{10}
\nn\\
&+ \frac{d (-220 + 309 d - 68 d^2 - 38 d^3 + 8 d^4 + d^5)}{4 (d-2) (2d-1) (-22 + 26 d - 9 d^2 + d^3)} c_{11}
\nn\\
&-  \frac{(72 - 77 d -  d^2 + 11 d^3 + d^4)}{6 (d-2) (d-1) (2d-1)}    c_{13} - \frac{1}{d-1} c_{14} -  \frac{ (60 - 47 d - 13 d^2 + 11 d^3 + d^4)}{6 (d-2) (d-1) (2d-1)}    c_{15}
\nn\\
&-  \frac{ (d + 7) (-12 + 5 d + d^2)}{6 (d-2) (2d-1)} c_{16} -  \frac{ d (d + 7) (-12 + 5 d + d^2)}{12 (d-2) (2d-1)} c_{18}- 2 (d-2) d   c_{26}
\end{align}

\begin{small}
\begin{align}
c_{23} &= -  \frac{(7104 - 35088 d + 67484 d^2 - 52018 d^3 + 8628 d^4 + 11187 d^5 - 7810 d^6 + 2236 d^7 - 288 d^8 + 3 d^9 + 2 d^{10})}{24 (d-2)^2 (d-1) (2d-1) (3d - 2) (-22 + 26 d - 9 d^2 + d^3)} c_1
\nn\\
&-  \frac{(13824 - 65760 d + 107232 d^2 - 80268 d^3 + 21812 d^4 + 7485 d^5 - 7491 d^6 + 2305 d^7 - 295 d^8 + 2 d^9 + 2 d^{10})}{12 (d-2)^2 (d-1) (2d-1) (3d - 2) (-22 + 26 d - 9 d^2 + d^3)} c_2
\nn\\
&-  \frac{(-6144 + 28032 d - 37964 d^2 + 27334 d^3 - 11888 d^4 + 2799 d^5 - 64 d^6 - 98 d^7 + 8 d^8 + d^9)}{24 (d-2)^2 (d-1) (2d-1) (3d - 2) (-22 + 26 d - 9 d^2 + d^3)} c_3
\nn\\
&- \frac{(15168 - 73200 d + 120720 d^2 - 92556 d^3 + 26636 d^4 + 7372 d^5 - 8018 d^6 + 2449 d^7 - 302 d^8 + d^9 + 2 d^{10})}{6 (d-2)^2 (d-1) (2d-1) (3d - 2) (-22 + 26 d - 9 d^2 + d^3)} c_4
\nn\\
&-  \frac{(15168 - 73200 d + 120720 d^2 - 92556 d^3 + 26636 d^4 + 7372 d^5 - 8018 d^6 + 2449 d^7 - 302 d^8 + d^9 + 2 d^{10})}{3 (d-2)^2 (d-1) (2d-1) (3d - 2) (-22 + 26 d - 9 d^2 + d^3)}  c_5
\nn\\
&-  \frac{(6624 - 31560 d + 52724 d^2 - 39676 d^3 + 10258 d^4 + 4194 d^5 - 3873 d^6 + 1167 d^7 - 148 d^8 + d^9 + d^{10})}{3 (d-2)^2 (d-1) (2d-1) (3d - 2) (-22 + 26 d - 9 d^2 + d^3)} c_6 
\nn\\
&-  \frac{(-5856 + 7872 d - 1872 d^2 - 1008 d^3 + 630 d^4 - 128 d^5 + 5 d^6 + d^7)}{3 (d-2) (3d - 2) (-22 + 26 d - 9 d^2 + d^3)} c_7
\nn\\
&-  \frac{(-1856 + 8016 d - 10784 d^2 + 6260 d^3 - 618 d^4 - 1205 d^5 + 722 d^6 - 160 d^7 + 8 d^8 + d^9)}{16 (d-2)^2 (d-1) (2d-1) (3d - 2) (-22 + 26 d - 9 d^2 + d^3)} c_8 
\nn\\
& -  \frac{ (12096 - 55344 d + 88108 d^2 - 60318 d^3 + 9758 d^4 + 10317 d^5 - 6659 d^6 + 1620 d^7 - 152 d^8 - 3 d^9 + d^{10})}{8 (d-2)^2 (d-1) (2d-1) (3d - 2) (-22 + 26 d - 9 d^2 + d^3)} c_9
\nn\\
&-  \frac{(20352 - 98400 d + 177616 d^2 - 139276 d^3 + 34518 d^4 + 16968 d^5 - 14063 d^6 + 3850 d^7 - 410 d^8 - 6 d^9 + 3 d^{10})}{48 (d-2)^2 (d-1) (2d-1) (3d - 2) (-22 + 26 d - 9 d^2 + d^3)} c_{10}
\nn\\
&-  \frac{(672 - 2232 d + 2438 d^2 - 1299 d^3 + 448 d^4 - 86 d^5 + 2 d^6 + d^7)}{8 (d-2)^2 (2d-1) (-22 + 26 d - 9 d^2 + d^3)} c_{11}  
\nn\\
&+ \frac{ (-96 + 168 d - 29 d^2 - 31 d^3 + 5 d^4 + d^5)}{12 (d-2)^2 (d-1) (2d-1)} c_{13}
\nn\\
&+\frac{ (d - 4) d}{2 (d-2)^2 (d-1)} c_{14} + \frac{(-120 + 222 d - 41 d^2 - 31 d^3 + 5 d^4 + d^5)}{12 (d-2)^2 (d-1) (2d-1)} c_{15}
\nn\\
&+ \frac{(72 - 42 d - 25 d^2 + 6 d^3 + d^4)}{12 (d-2)^2 (2d-1)} c_{16} + + \frac{ (96 - 168 d + 66 d^2 - 49 d^3 + 6 d^4 + d^5)}{24 (d-2)^2 (2d-1)} c_{18}
\nn\\
& + (12 - 6 d + d^2)   c_{26}
\end{align}
\end{small}

\begin{align}
c_{24} &= -  \frac{ (1768 - 5050 d + 5912 d^2 - 3707 d^3 + 1391 d^4 - 285 d^5 + 17 d^6 + 2 d^7)}{3 (d-2) (2d-1) (3d - 2) (-22 + 26 d - 9 d^2 + d^3)} c_{1}
\nn\\
&-  \frac{2  (-560 - 376 d + 2678 d^2 - 2965 d^3 + 1408 d^4 - 299 d^5 + 16 d^6 + 2 d^7)}{3 (d-2) (2d-1) (3d - 2) (-22 + 26 d - 9 d^2 + d^3)} c_2
\nn\\
&-  \frac{(1696 - 3274 d + 2116 d^2 - 371 d^3 - 63 d^4 + 15 d^5 + d^6)}{3 (d-2) (2d-1) (3d - 2) (-22 + 26 d - 9 d^2 + d^3)} c_3
\nn\\
&-  \frac{4 (-920 + 196 d + 2624 d^2 - 3272 d^3 + 1548 d^4 - 313 d^5 + 15 d^6 + 2 d^7)}{3 (d-2) (2d-1) (3d - 2) (-22 + 26 d - 9 d^2 + d^3)} c_4
\nn\\
&-  \frac{8  (-920 + 196 d + 2624 d^2 - 3272 d^3 + 1548 d^4 - 313 d^5 + 15 d^6 + 2 d^7)}{3 (d-2) (2d-1) (3d - 2) (-22 + 26 d - 9 d^2 + d^3)}   c_5
\nn\\
&-  \frac{8 (36 - 888 d + 1898 d^2 - 1668 d^3 + 727 d^4 - 150 d^5 + 8 d^6 + d^7)}{3 (d-2) (2d-1) (3d - 2) (-22 + 26 d - 9 d^2 + d^3)} c_6 
\nn\\
&-  \frac{8 (d-2) (-168 + 190 d - 88 d^2 + 13 d^3 + d^4)}{3 (3d - 2) (-22 + 26 d - 9 d^2 + d^3)} c_7
\nn\\
& -  \frac{ (120 + 44 d - 416 d^2 + 417 d^3 - 149 d^4 + 15 d^5 + d^6)}{2 (d-2) (2d-1) (3d - 2) (-22 + 26 d - 9 d^2 + d^3)} c_8
 \nn\\
&-  \frac{ (632 - 3322 d + 5142 d^2 - 3529 d^3 + 1162 d^4 - 170 d^5 + 4 d^6 + d^7)}{(d-2) (2d-1) (3d - 2) (-22 + 26 d - 9 d^2 + d^3)} c_9
\nn\\ 
&-  \frac{(2544 - 8320 d + 10510 d^2 - 6760 d^3 + 2395 d^4 - 419 d^5 + 15 d^6 + 3 d^7)}{6 (d-2) (2d-1) (3d - 2) (-22 + 26 d - 9 d^2 + d^3)} c_{10}
\nn\\
&-  \frac{(d-1) (-4 + 55 d - 43 d^2 + 5 d^3 + d^4)}{(d-2) (2d-1) (-22 + 26 d - 9 d^2 + d^3)} c_{11}
\nn\\
& + \frac{2  (-15 + 8 d + d^2)}{3 (d-2) (2d-1)} c_{13} +  \frac{2  (-15 + 8 d + d^2)}{3 (d-2) (2d-1)} c_{15}
\nn\\
&+ \frac{2  (-15 + 8 d + d^2)}{3 (d-2) (2d-1)} c_{16} + \frac{d (-15 + 8 d + d^2)}{3 (d-2) (2d-1)} c_{18}   + 8 (d-2) c_{26}      
\end{align}

\begin{align}
c_{25} &= + \frac{(3600 - 16408 d + 31034 d^2 - 30162 d^3 + 16863 d^4 - 5794 d^5 + 1228 d^6 - 114 d^7 - 9 d^8 + 2 d^9)}{12 (d-2)^2 (d-1) (2d-1) (3d - 2) (-22 + 26 d - 9 d^2 + d^3)} c_1
\nn\\
&+ \frac{ (912 - 7080 d + 15960 d^2 - 18454 d^3 + 12579 d^4 - 5229 d^5 + 1243 d^6 - 115 d^7 - 10 d^8 + 2 d^9)}{6 (d-2)^2 (d-1) (2d-1) (3d - 2) (-22 + 26 d - 9 d^2 + d^3)} c_2
\nn\\
&+ \frac{(1824 - 6056 d + 9634 d^2 - 7274 d^3 + 2439 d^4 - 184 d^5 - 50 d^6 + 2 d^7 + d^8)}{12 (d-2)^2 (d-1) (2d-1) (3d - 2) (-22 + 26 d - 9 d^2 + d^3)} c_3
\nn\\
&+ \frac{ (480 - 5880 d + 14820 d^2 - 18556 d^3 + 13312 d^4 - 5672 d^5 + 1333 d^6 - 116 d^7 - 11 d^8 + 2 d^9)}{3 (d-2)^2 (d-1) (2d-1) (3d - 2) (-22 + 26 d - 9 d^2 + d^3)} c_4
\nn\\
&+ \frac{2 (480 - 5880 d + 14820 d^2 - 18556 d^3 + 13312 d^4 - 5672 d^5 + 1333 d^6 - 116 d^7 - 11 d^8 + 2 d^9)}{3 (d-2)^2 (d-1) (2d-1) (3d - 2) (-22 + 26 d - 9 d^2 + d^3)} c_5
\nn\\
&+ \frac{2 (888 - 5176 d + 10700 d^2 - 11444 d^3 + 7212 d^4 - 2805 d^5 + 639 d^6 - 58 d^7 - 5 d^8 + d^9)}{3 (d-2)^2 (d-1) (2d-1) (3d - 2) (-22 + 26 d - 9 d^2 + d^3)} c_6    
\nn\\
&+ \frac{2 (-1248 + 1968 d - 1116 d^2 + 366 d^3 - 62 d^4 -  d^5 + d^6)}{3 (d-2) (3d - 2) (-22 + 26 d - 9 d^2 + d^3)} c_7
\nn\\
&+ \frac{(-32 + 600 d - 1476 d^2 + 1936 d^3 - 1469 d^4 + 614 d^5 - 112 d^6 + 2 d^7 + d^8)}{8 (d-2)^2 (d-1) (2d-1) (3d - 2) (-22 + 26 d - 9 d^2 + d^3)} c_8
\nn\\
&+ \frac{(2448 - 12968 d + 25506 d^2 - 25744 d^3 + 14727 d^4 - 4859 d^5 + 840 d^6 - 38 d^7 - 9 d^8 + d^9)}{4 (d-2)^2 (d-1) (2d-1) (3d - 2) (-22 + 26 d - 9 d^2 + d^3)} c_9
\nn\\
&+ \frac{(6432 - 30800 d + 58832 d^2 - 57438 d^3 + 31686 d^4 - 10223 d^5 + 1810 d^6 - 86 d^7 - 24 d^8 + 3 d^9)}{24 (d-2)^2 (d-1) (2d-1) (3d - 2) (-22 + 26 d - 9 d^2 + d^3)} c_{10}
\nn\\
&+ \frac{(d-3) (-16 + 12 d + 37 d^2 - 29 d^3 -  d^4 + d^5)}{4 (d-2)^2 (2d-1) (-22 + 26 d - 9 d^2 + d^3)}  c_{11}  
\nn\\
&-  \frac{ (-33 + 61 d - 25 d^2 -  d^3 + d^4)}{6 (d-2)^2 (d-1) (2d-1)} c_{13}
\nn\\
&+ \frac{ d}{2 (d-2)^2 (d-1)} c_{14} -  \frac{ (-36 + 67 d - 25 d^2 -  d^3 + d^4)}{6 (d-2)^2 (d-1) (2d-1)} c_{15} -  \frac{(30 - 25 d + d^3)}{6 (d-2)^2 (2d-1)}  c_{16}
\nn\\
&-  \frac{ (d-1) d (-36 + d + d^2)}{12 (d-2)^2 (2d-1)}c_{18}-2  d  c_{26}  
\end{align}

\section{Quasi-topological Lagrangian densities}
\label{app:QT_lags}

In this appendix we provide a list of the explicit forms of the quasi-topological Lagrangian densities.

\begin{align}
\mathcal{Z}_d^{(1)} &= 16 (d - 2) (244 - 451 d + 306 d^2 - 91 d^3 + 10 d^4) R_{a}{}^{c} R^{ab} R_{b}{}^{d} R_{cd} 
\nn\\
&- 64 (d - 2) (7 - 5 d + d^2) (14 - 14 d + 3 d^2) R_{a}{}^{c} R^{ab} R_{bc} R 
\nn\\
&+ 8 (-388 + 931 d - 856 d^2 + 379 d^3 - 82 d^4 + 7 d^5) R_{ab} R^{ab} R^2 
\nn\\
&+ (-980 + 1683 d - 1060 d^2 + 302 d^3 - 36 d^4 + d^5) R^4
\nn\\
& - 32 (d - 4)^2 (d - 2)^2 (14 - 14 d + 3 d^2) R^{ab} R^{cd} R R_{acbd}
\nn\\
&+ 2 (2764 - 6289 d + 5788 d^2 - 2776 d^3 + 736 d^4 - 103 d^5 + 6 d^6) R^2 R_{abcd} R^{abcd}
\nn\\
& + 64 (d - 3) (d - 2)^2 (-58 + 75 d - 30 d^2 + 4 d^3) R_{a}{}^{c} R^{ab} R^{de} R_{bdce}
\nn\\
& - 48 (d - 3) (d - 2) (4 - 31 d + 37 d^2 - 15 d^3 + 2 d^4) R^{ab} R^{cd} R_{ac}{}^{ef} R_{bdef}
\nn\\
&+ 16 (d - 2)^3 (274 - 389 d + 183 d^2 - 34 d^3 + 2 d^4) R^{ab} R^{cd} R_{a}{}^{e}{}_{b}{}^{f} R_{cedf} 
\nn\\
&- 4 (d - 4) (118 - 596 d + 876 d^2 - 581 d^3 + 195 d^4 - 32 d^5 + 2 d^6) R_{ab} R^{ab} R_{cdef} R^{cdef} 
\nn\\
&+ 16 (d - 4) (d - 3) (d - 2) (d - 1) (14 - 14 d + 3 d^2) R^{ab} R_{a}{}^{cde} R_{bc}{}^{fh} R_{defh}
\nn\\
& -  (d - 2) (1108 - 2723 d + 2639 d^2 - 1224 d^3 
\nn\\
&+ 235 d^4 + 10 d^5 - 10 d^6 + d^7) R_{ab}{}^{ef} R^{abcd} R_{cd}{}^{hi} R_{efhi} + 8 (d - 2) (860 - 2113 d
\nn\\
&  + 1959 d^2 - 810 d^3 + 102 d^4 + 30 d^5 - 11 d^6 + d^7) R_{a}{}^{e}{}_{c}{}^{f} R^{abcd} R_{b}{}^{h}{}_{d}{}^{i} R_{ehfi} 
\nn\\
&+ (-1292 + 2929 d - 2741 d^2 + 1527 d^3 - 684 d^4 + 276 d^5 - 82 d^6 
\nn\\
& + 14 d^7 -  d^8) R_{abcd} R^{abcd} R_{efhi} R^{efhi}
\end{align}

\beqa
\mathcal{Z}_d^{(2)}&=& \frac{1}{(d - 4) (d - 2)^3 (3d - 4) (11 - 6 d + d^2) (-4 + 14 d - 7 d^2 + d^3)(-22 + 26 d - 9 d^2 + d^3)}\times\nonumber\\
&&\left.\times \big[(d-4) (d^3-9 d^2+26 d-22) (d-1) (2 d^8-36 d^7+264 d^6-969 d^5+1486 d^4+1289 \right.\nonumber\\
&&\left.\times d^3-8530 d^2+11948 d-5632) R^{ab} R_{ab} R^{cd} R_{cd} -(d-2)(-22 + 26 d - 9 d^2 + d^3)(3840\right.\nonumber\\
&&\left. - 9872 d + 13772 d^2 - 12446 d^3 + 6133 d^4- 795 d^5 - 639 d^6 + 327 d^7 - 60 d^8 + 4 d^9) R_{a}{}^{c} R^{ab}\right.\nonumber\\
&&\left. \times R_{b}{}^{d} R_{cd}
+ (d - 1)(d - 4)(-22 + 26 d- 9 d^2 + d^3) (-5632 + 11948 d - 8530 d^2 + 1289 d^3 \right.\nonumber\\
&&\left.+ 1486 d^4 - 969 d^5 + 264 d^6- 36 d^7 + 2 d^8)R_{a}{}^{c} R^{ab} R_{b}{}^{d} R_{cd}
+ \frac{4}{3}(d-2) (92672 - 459640 d\right.\nonumber\\
&&\left. + 851460 d^2 - 741570 d^3 + 245584 d^4 + 91339 d^5 - 122856 d^6 + 51524 d^7 - 10130 d^8 + 451 \right.\nonumber\\
&&\left. \times d^9+ 192 d^{10} - 36 d^{11} + 2 d^{12})  R_{a}{}^{c} R^{ab} R_{bc} R - (-19968 + 129856 d - 351080 d^2+ 486664\right.\nonumber\\
&&\left.\times d^3 - 350864 d^4 + 91452 d^5  + 48784 d^6 - 50566 d^7 + 18113 d^8 - 2536 d^9- 243 d^{10} + 143 d^{11}\right.\nonumber\\
&&\left.  - 20 d^{12} + d^{13})R_{ab} R^{ab} R^2 + \frac{1}{24}(385024 - 1950016 d+ 3753760 d^2 - 3555864 d^3 + 1582172 \right.\nonumber\\
&&\left. \times d^4 - 4394 d^5- 370858 d^6 + 206017 d^7- 59436 d^8 + 10909 d^9 - 1522 d^{10} + 195 d^{11} - 20 d^{12} \right.\nonumber\\
&&\left.+ d^{13}) R^4 + 4(d-2)^2(-36480+ 97652 d - 90614 d^2 + 16524 d^3 + 30278 d^4 - 24508 d^5 \right.\nonumber\\
&&\left.+ 6916 d^6 + 152 d^7 - 625 d^8+ 173 d^9 - 21 d^{10} + d^{11}) R R^{ab} R^{cd} R_{acbd} + \frac{1}{4}(d-2)(-328704\right.\nonumber\\
&&\left. + 1158096 d- 1701488 d^2 + 1286084 d^3 - 430702 d^4 - 82374 d^5 + 157229 d^6 - 79874 d^7\right.\nonumber\\
&&\left.+ 23397 d^8 - 4346 d^9 + 507 d^{10} - 34 d^{11} + d^{12}) R^2 R_{abcd} R^{abcd}- 2 (d-2)^3 (-28032 + 87822 d \right.\nonumber\\
&&\left.- 112640 d^2 + 71315 d^3 - 16827 d^4 - 6654 d^5+ 6558 d^6 - 2329 d^7 + 447 d^8 - 46 d^9 + 2 d^{10})  R^{ab} \right.\nonumber\\
&&\left.\times R R_{a}{}^{cde} R_{bcde}- 8 (d-2)^2(-22 + 26 d - 9 d^2 + d^3)(-64 + 1592 d - 2909 d^2 + 1743 d^3- 58 d^4\right.\nonumber\\
&&\left. - 371 d^5 + 167 d^6 - 30 d^7 + 2 d^8) R_{a}{}^{c} R^{ab} R^{de} R_{bdce}+ (d-2)^3(-22 + 26 d - 9 d^2 + d^3)(1024 \right.\nonumber\\
&&\left.- 3308 d + 2725 d^2 + 210 d^3- 1190 d^4 + 570 d^5 - 111 d^6 + 8 d^7) R^{ab} R^{cd} R_{ac}{}^{ef} R_{bdef} + \frac{1}{3}(d-2)^3\right.\nonumber\\
&&\left.\times(1792 - 3743 d+ 2678 d^2 - 531 d^3 - 247 d^4 + 150 d^5 - 29 d^6 + 2 d^7) (-4 + 14 d - 7 d^2 + d^3)\right.\nonumber\\
&&\left.\times  R R_{ab}{}^{ef} R^{abcd} R_{cdef} + \frac{1}{2} (d-2)(-22 + 26 d - 9 d^2 + d^3) (9216- 31760 d + 41152 d^2 - 22702 \right.\nonumber\\
&&\left.\times d^3 + 914 d^4 + 5611 d^5 - 3201 d^6 + 839 d^7- 111 d^8 + 6 d^9) R_{ab} R^{ab} R_{cdef} R^{cdef}
\big]\right.\nonumber\\
&&\left.-  \frac{2 (-8 - 23 d + 39 d^2 - 16 d^3 + 2 d^4) }{(d - 4) (3d - 4) (11 - 6 d + d^2)}  R^{ab} R_{a}{}^{cde} R_{bc}{}^{fh} R_{defh} + R_{a}{}^{e}{}_{c}{}^{f} R^{abcd} R_{b}{}^{h}{}_{e}{}^{j} R_{dhfj}\right.
\eeqa

\begin{align}
\mathcal{Z}_d^{(3)} &=  \frac{1}{12 (-4 + d) (-2 + d)^3 (-4 + 3 d) (11 - 6 d + d^2) (-22 + 26 d - 9 d^2 + d^3) (-4 + 14 d - 7 d^2 + d^3)} \times
\nn\\
&\times \big[ 
-24 (-2 + d) (-22 + 26 d - 9 d^2 + d^3) (-3408 + 9452 d - 13070 d^2 + 12869 d^3 
\nn\\
&- 9751 d^4 + 5409 d^5 - 2053 d^6 + 496 d^7 - 68 d^8 + 4 d^9) R_{a}{}^{c} R^{ab} R_{b}{}^{d} R_{cd} 
\nn\\
&+ 24 (-4 + d) (-3 + d) (-22 + 26 d - 9 d^2 + d^3) (1716 - 5894 d + 8839 d^2 - 7538 d^3 
\nn\\
&+ 4008 d^4 - 1364 d^5 + 291 d^6 - 36 d^7 + 2 d^8) R_{ab} R^{ab} R_{cd} R^{cd} + 32 (-2 + d) (-71704 
\nn\\
&+ 400996 d - 956122 d^2 + 1301340 d^3 - 1128581 d^4 + 652069 d^5 - 251257 d^6 
\nn\\
&+ 60923 d^7 - 7184 d^8 - 444 d^9 + 290 d^{10} - 40 d^{11} + 2 d^{12}) R_{a}{}^{c} R^{ab} R_{bc} R - 24 (15680 
\nn\\
&- 106664 d + 323592 d^2 - 568168 d^3 + 638164 d^4 - 479674 d^5 + 243364 d^6 - 80096 d^7 
\nn\\
&+ 14246 d^8 + 229 d^9 - 800 d^{10} + 196 d^{11} - 22 d^{12} + d^{13}) R_{ab} R^{ab} R^2 + (-302144 
\nn\\
&+ 1720608 d - 4189176 d^2 + 5863660 d^3 - 5304058 d^4 + 3284002 d^5 - 1431861 d^6 
\nn\\
&+ 445160 d^7 - 99552 d^8 + 16457 d^9 - 2171 d^{10} + 248 d^{11} - 22 d^{12} + d^{13}) R^4 
\nn\\
&+ 48 (-2 + d)^2 (70472 - 240892 d + 359520 d^2 - 299804 d^3 + 143976 d^4 - 30793 d^5 
\nn\\
&- 6594 d^6 + 7094 d^7 - 2428 d^8 + 453 d^9 - 46 d^{10} + 2 d^{11}) R^{ab} R^{cd} R R_{acbd} 
\nn\\
&+ 6 (-2 + d) (316048 - 1340112 d + 2613908 d^2 - 3095774 d^3 + 2474698 d^4 
\nn\\
&- 1403521 d^5 + 577724 d^6 - 173518 d^7 + 37673 d^8 - 5759 d^9 + 588 d^{10} 
\nn\\
&- 36 d^{11} + d^{12}) R^2 R_{abcd} R^{abcd} - 48 (-2 + d)^3 (25270 - 92828 d + 156501 d^2 
\nn\\
&- 158736 d^3 + 107067 d^4 - 50145 d^5 + 16490 d^6 - 3749 d^7 + 562 d^8 - 50 d^9 
\nn\\
&+ 2 d^{10}) R^{ab} R R_{a}{}^{cde} R_{bcde} - 192 (-2 + d)^2 (-22 + 26 d - 9 d^2 + d^3) (-40 - 1069 d 
\nn\\
&+ 3085 d^2 - 3689 d^3 + 2463 d^4 - 1001 d^5 + 247 d^6 - 34 d^7 + 2 d^8) R_{a}{}^{c} R^{ab} R^{de} R_{bdce} 
\nn\\
&+ 24 (-2 + d)^3 (-1 + d) (-22 + 26 d - 9 d^2 + d^3) (988 - 3197 d + 3661 d^2 - 2141 d^3 
\nn\\
&+ 689 d^4 - 116 d^5 + 8 d^6) R^{ab} R^{cd} R_{ac}{}^{ef} R_{bdef} + 4 (-2 + d)^3 (-4 + 14 d 
\nn\\
&- 7 d^2 + d^3) (-1654 + 4528 d - 5595 d^2 + 4003 d^3 - 1751 d^4 + 459 d^5 - 66 d^6 
\nn\\
&+ 4 d^7) R R_{ab}{}^{ef} R^{abcd} R_{cdef} + 12 (-2 + d) (-22 + 26 d - 9 d^2 + d^3) (-8144 + 33248 d 
\nn\\
&- 60534 d^2 + 64054 d^3 - 43371 d^4 + 19504 d^5 - 5824 d^6 + 1112 d^7 
\nn\\
&- 123 d^8 + 6 d^9) R_{ab} R^{ab} R_{cdef} R^{cdef} - 24 (-2 + d)^3 (-22 + 26 d - 9 d^2 + d^3) (-4 
\nn\\
&+ 14 d - 7 d^2 + d^3) (122 - 207 d + 130 d^2 - 37 d^3 + 4 d^4) R^{ab} R_{a}{}^{cde} R_{bc}{}^{fh} R_{defh} 
\big]
\nn\\
& + R_{a}{}^{e}{}_{c}{}^{f} R^{abcd} R_{b}{}^{h}{}_{d}{}^{j} R_{ehfj}
\end{align}

\begin{align}
\mathcal{Z}_d^{(4)} &= \frac{1}{12 (-2 + d)^2 (-4 + 3 d) (11 - 6 d + d^2) (-22 + 26 d - 9 d^2 + d^3) (-4 + 14 d - 7 d^2 + d^3)} \times
\nn\\
&\times \big[ 
-48 (-2 + d) (-22 + 26 d - 9 d^2 + d^3) (136 - 230 d + 271 d^2 - 248 d^3 
\nn\\
&+ 119 d^4 - 26 d^5 + 2 d^6) R_{a}{}^{c} R^{ab} R_{b}{}^{d} R_{cd} + 48 (-1 + d) (-22 + 26 d - 9 d^2 + d^3) (968 
\nn\\
&- 2030 d + 1645 d^2 - 689 d^3 + 161 d^4 - 20 d^5 + d^6) R_{ab} R^{ab} R_{cd} R^{cd} 
\nn\\
&+ 16 (-2 + d) (16144 - 75888 d + 132572 d^2 - 115700 d^3 + 54596 d^4 - 13179 d^5 
\nn\\
&+ 902 d^6 + 277 d^7 - 64 d^8 + 4 d^9) R_{a}{}^{c} R^{ab} R_{bc} R - 24 (-1520 + 9492 d - 24910 d^2 
\nn\\
&+ 33458 d^3 - 24719 d^4 + 9944 d^5 - 1768 d^6 - 118 d^7 + 110 d^8 - 18 d^9 + d^{10}) R_{ab} R^{ab} R^2 
\nn\\
&+ (33280 - 158960 d + 285656 d^2 - 263172 d^3 + 139206 d^4 - 44518 d^5 + 8963 d^6 
\nn\\
&- 1272 d^7 + 162 d^8 - 18 d^9 + d^{10}) R^4 + 96 (-2 + d)^2 (-3112 + 7497 d - 6676 d^2 
\nn\\
&+ 2251 d^3 + 265 d^4 - 441 d^5 + 138 d^6 - 19 d^7 + d^8) R^{ab} R^{cd} R R_{acbd} 
\nn\\
&+ 6 (-2 + d) (-28000 + 90828 d - 127196 d^2 + 100724 d^3 - 49778 d^4 + 15961 d^5 
\nn\\
&- 3326 d^6 + 434 d^7 - 32 d^8 + d^9) R^2 R_{abcd} R^{abcd} - 24 (-2 + d)^2 (9280 - 30290 d 
\nn\\
&+ 42690 d^2 - 33711 d^3 + 16264 d^4 - 4901 d^5 + 900 d^6 - 92 d^7 + 4 d^8) R^{ab} R R_{a}{}^{cde} R_{bcde} 
\nn\\
&- 48 (-2 + d)^2 (-22 + 26 d - 9 d^2 + d^3) (-24 + 538 d - 817 d^2 + 444 d^3 - 101 d^4 
\nn\\
&+ 8 d^5) R_{a}{}^{c} R^{ab} R^{de} R_{bdce} + 12 (-2 + d)^2 (-22 + 26 d - 9 d^2 + d^3) (-384 + 1396 d 
\nn\\
&- 1615 d^2 + 781 d^3 - 167 d^4 + 13 d^5) R^{ab} R^{cd} R_{ac}{}^{ef} R_{bdef} + 4 (-2 + d)^2 (-4 + 14 d 
\nn\\
&- 7 d^2 + d^3) (-496 + 1049 d - 844 d^2 + 321 d^3 - 58 d^4 + 4 d^5) R R_{ab}{}^{ef} R^{abcd} R_{cdef} 
\nn\\
&+ 12 (-4 + d) (-2 + d) (-1 + d) (-22 + 26 d - 9 d^2 + d^3) (194 - 370 d + 237 d^2 
\nn\\
&- 63 d^3 + 6 d^4) R_{ab} R^{ab} R_{cdef} R^{cdef} - 12 (-4 + d) (-2 + d)^2 (-7 + 5 d) (-22 + 26 d 
\nn\\
&- 9 d^2 + d^3) (-4 + 14 d - 7 d^2 + d^3) R^{ab} R_{a}{}^{cde} R_{bc}{}^{fh} R_{defh} \big] + R_{ab}{}^{ef} R^{abcd} R_{c}{}^{h}{}_{e}{}^{j} R_{dhfj}
\end{align}

\begin{align}
\mathcal{Z}_d^{(5)} &= \frac{1}{6 (-4 + d) (-2 + d)^3 (-4 + 3 d) (11 - 6 d + d^2) (-22 + 26 d - 9 d^2 + d^3) (-4 + 14 d - 7 d^2 + d^3)} \times
\nn\\
&\times \big[ 
-48 (-2 + d) (-22 + 26 d - 9 d^2 + d^3) (-144 + 188 d - 1374 d^2 + 4021 d^3 
\nn\\
&- 5045 d^4 + 3387 d^5 - 1325 d^6 + 304 d^7 - 38 d^8 + 2 d^9) R_{a}{}^{c} R^{ab} R_{b}{}^{d} R_{cd} 
\nn\\
&+ 48 (-4 + d) (-22 + 26 d - 9 d^2 + d^3) (-1804 + 9398 d - 18611 d^2 + 19639 d^3 
\nn\\
&- 12568 d^4 + 5162 d^5 - 1380 d^6 + 234 d^7 - 23 d^8 + d^9) R_{ab} R^{ab} R_{cd} R^{cd} 
\nn\\
&+ 64 (-2 + d) (-39608 + 263092 d - 689234 d^2 + 982248 d^3 - 860735 d^4 
\nn\\
&+ 487978 d^5 - 179717 d^6 + 40454 d^7 - 4102 d^8 - 390 d^9 + 179 d^{10} 
\nn\\
&- 22 d^{11} + d^{12}) R_{a}{}^{c} R^{ab} R_{bc} R - 24 (4736 - 58896 d + 273136 d^2 - 618416 d^3
\nn\\
& + 804696 d^4 - 652724 d^5 + 339308 d^6 - 109530 d^7 + 18000 d^8 + 735 d^9 - 1064 d^{10} 
\nn\\
&+ 234 d^{11} - 24 d^{12} + d^{13}) R_{ab} R^{ab} R^2 + (-318080 + 2132288 d - 5677552 d^2 
\nn\\
&+ 8307160 d^3 - 7598852 d^4 + 4634276 d^5 - 1950058 d^6 + 577374 d^7 - 122648 d^8 
\nn\\
&+ 19523 d^9 - 2542 d^{10} + 286 d^{11} - 24 d^{12} + d^{13}) R^4 + 96 (-2 + d)^2 (17960 - 85612 d 
\nn\\
&+ 156176 d^2 - 144780 d^3 + 69622 d^4 - 10116 d^5 - 7511 d^6 + 5207 d^7 - 1574 d^8 
\nn\\
&+ 268 d^9 - 25 d^{10} + d^{11}) R^{ab} R^{cd} R R_{acbd} + 6 (-2 + d) (178336 - 1018016 d + 2405768 d^2 
\nn\\
&- 3225004 d^3 + 2781796 d^4 - 1644450 d^5 + 687962 d^6 - 206196 d^7 + 44081 d^8 
\nn\\
&- 6568 d^9 + 648 d^{10} - 38 d^{11} + d^{12}) R^2 R_{abcd} R^{abcd} - 96 (-2 + d)^3 (7550 - 39268 d 
\nn\\
&+ 81391 d^2 - 93331 d^3 + 67198 d^4 - 32176 d^5 + 10461 d^6 - 2292 d^7 + 325 d^8 
\nn\\
&- 27 d^9 + d^{10}) R^{ab} R R_{a}{}^{cde} R_{bcde} - 384 (-2 + d)^2 (-1 + d) (-22 + 26 d - 9 d^2
\nn\\
& + d^3) (-152 + 929 d - 1562 d^2 + 1239 d^3 - 542 d^4 + 135 d^5 - 18 d^6 + d^7) R_{a}{}^{c} R^{ab} R^{de} R_{bdce}
\nn\\
& + 48 (-2 + d)^3 (-22 + 26 d - 9 d^2 + d^3) (-716 + 3557 d - 5760 d^2 + 4566 d^3 - 2022 d^4 
\nn\\
&+ 513 d^5 - 70 d^6 + 4 d^7) R^{ab} R^{cd} R_{ac}{}^{ef} R_{bdef} + 8 (-2 + d)^3 (-4 + 14 d - 7 d^2 + d^3) (-878 
\nn\\
&+ 3064 d - 4182 d^2 + 2976 d^3 - 1215 d^4 + 288 d^5 - 37 d^6 + 2 d^7) R R_{ab}{}^{ef} R^{abcd} R_{cdef} 
\nn\\
&+ 24 (-2 + d) (-22 + 26 d - 9 d^2 + d^3) (-2704 + 14944 d - 32382 d^2 + 37746 d^3 
\nn\\
&- 26667 d^4 + 12018 d^5 - 3491 d^6 + 635 d^7 - 66 d^8 + 3 d^9) R_{ab} R^{ab} R_{cdef} R^{cdef} 
\nn\\
&- 48 (-2 + d)^3 (-22 + 26 d - 9 d^2 + d^3) (-4 + 14 d - 7 d^2 + d^3) (82 - 139 d + 82 d^2 
\nn\\
&- 21 d^3 + 2 d^4) R^{ab} R_{a}{}^{cde} R_{bc}{}^{fh} R_{defh} \big] + R_{ab}{}^{ef} R^{abcd} R_{ce}{}^{hj} R_{dfhj}
\end{align}

\begin{align}
\mathcal{Z}_d^{(6)} &= \frac{1}{3 (-4 + d) (-2 + d)^3 (-4 + 3 d) (11 - 6 d + d^2) (-22 + 26 d - 9 d^2 + d^3) (-4 + 14 d - 7 d^2 + d^3)} \times
\nn\\
&\times \big[ 
-48 (-2 + d) (-22 + 26 d - 9 d^2 + d^3) (-144 + 188 d - 1374 d^2 + 4021 d^3 - 5045 d^4 
\nn\\
&+ 3387 d^5 - 1325 d^6 + 304 d^7 - 38 d^8 + 2 d^9) R_{a}{}^{c} R^{ab} R_{b}{}^{d} R_{cd} + 48 (-4 + d) (-22 
\nn\\
&+ 26 d - 9 d^2 + d^3) (-1804 + 9398 d - 18611 d^2 + 19639 d^3 - 12568 d^4 + 5162 d^5 
\nn\\
&- 1380 d^6 + 234 d^7 - 23 d^8 + d^9) R_{ab} R^{ab} R_{cd} R^{cd} + 64 (-2 + d) (-39608 + 263092 d 
\nn\\
&- 689234 d^2 + 982248 d^3 - 860735 d^4 + 487978 d^5 - 179717 d^6 + 40454 d^7 - 4102 d^8 
\nn\\
&- 390 d^9 + 179 d^{10} - 22 d^{11} + d^{12}) R_{a}{}^{c} R^{ab} R_{bc} R - 24 (4736 - 58896 d + 273136 d^2 
\nn\\
&- 618416 d^3 + 804696 d^4 - 652724 d^5 + 339308 d^6 - 109530 d^7 + 18000 d^8 + 735 d^9 
\nn\\
&- 1064 d^{10} + 234 d^{11} - 24 d^{12} + d^{13}) R_{ab} R^{ab} R^2 + (-318080 + 2132288 d - 5677552 d^2 
\nn\\
&+ 8307160 d^3 - 7598852 d^4 + 4634276 d^5 - 1950058 d^6 + 577374 d^7 - 122648 d^8 + 19523 d^9 
\nn\\
&- 2542 d^{10} + 286 d^{11} - 24 d^{12} + d^{13}) R^4 + 96 (-2 + d)^2 (17960 - 85612 d + 156176 d^2 
\nn\\
&- 144780 d^3 + 69622 d^4 - 10116 d^5 - 7511 d^6 + 5207 d^7 - 1574 d^8 + 268 d^9 
\nn\\
&- 25 d^{10} + d^{11}) R^{ab} R^{cd} R R_{acbd} + 6 (-2 + d) (178336 - 1018016 d + 2405768 d^2 
\nn\\
&- 3225004 d^3 + 2781796 d^4 - 1644450 d^5 + 687962 d^6 - 206196 d^7 + 44081 d^8 
\nn\\
&- 6568 d^9 + 648 d^{10} - 38 d^{11} + d^{12}) R^2 R_{abcd} R^{abcd} - 96 (-2 + d)^3 (7550 - 39268 d 
\nn\\
&+ 81391 d^2 - 93331 d^3 + 67198 d^4 - 32176 d^5 + 10461 d^6 - 2292 d^7 + 325 d^8 
\nn\\
&- 27 d^9 + d^{10}) R^{ab} R R_{a}{}^{cde} R_{bcde} - 384 (-2 + d)^2 (-1 + d) (-22 + 26 d - 9 d^2 
\nn\\
&+ d^3) (-152 + 929 d - 1562 d^2 + 1239 d^3 - 542 d^4 + 135 d^5 - 18 d^6 + d^7) R_{a}{}^{c} R^{ab} R^{de} R_{bdce} 
\nn\\
&+ 48 (-2 + d)^3 (-22 + 26 d - 9 d^2 + d^3) (-716 + 3557 d - 5760 d^2 + 4566 d^3 - 2022 d^4 
\nn\\
&+ 513 d^5 - 70 d^6 + 4 d^7) R^{ab} R^{cd} R_{ac}{}^{ef} R_{bdef} + 8 (-2 + d)^3 (-4 + 14 d - 7 d^2 
\nn\\
&+ d^3) (-878 + 3064 d - 4182 d^2 + 2976 d^3 - 1215 d^4 + 288 d^5 - 37 d^6 + 2 d^7) R R_{ab}{}^{ef} R^{abcd} R_{cdef} 
\nn\\
&+ 24 (-2 + d) (-22 + 26 d - 9 d^2 + d^3) (-2704 + 14944 d - 32382 d^2 + 37746 d^3 - 26667 d^4 
\nn\\
&+ 12018 d^5 - 3491 d^6 + 635 d^7 - 66 d^8 + 3 d^9) R_{ab} R^{ab} R_{cdef} R^{cdef} 
\nn\\
&- 48 (-2 + d)^3 (-22 + 26 d - 9 d^2 + d^3) (-4 + 14 d - 7 d^2 + d^3) (82 - 139 d + 82 d^2 
\nn\\
&- 21 d^3 + 2 d^4) R^{ab} R_{a}{}^{cde} R_{bc}{}^{fh} R_{defh} \big] +  R_{ab}{}^{ef} R^{abcd} R_{cd}{}^{hj} R_{efhj}
\end{align}

\section{Generalized quasi-topological Lagrangian densities}
\label{app:GQT_lags}. 

Here we present the explicit forms of the quartet of generalized quasi-topological theories.

\begin{align}
\mathcal{S}_d^{(1)} &= 
\frac{1}{6 (d - 3)^2 (d - 2)^2 (d -1) d (11 - 6 d + d^2) (19 - 18 d + 3 d^2) (-22 + 26 d - 9 d^2 + d^3)}
\nn\\
&\times 
\big[
-2 (d - 2)^2 (675840 - 1895902 d + 2220384 d^2 - 1342691 d^3 + 370480 d^4 
\nn\\
&+ 36380 d^5 - 68962 d^6 + 27252 d^7 - 6100 d^8 + 862 d^9 - 74 d^{10} + 3 d^{11})  R_{a}{}^{c} R^{ab} R_{b}{}^{d} R_{cd}
\nn\\
& - 2 (1332480 - 3880512 d + 4484792 d^2 - 2299414 d^3 + 114412 d^4 + 452234 d^5 
\nn\\
&- 195096 d^6 - 509 d^7 + 26111 d^8 - 9952 d^9 + 1830 d^{10} - 175 d^{11} + 7 d^{12}) R_{ab} R^{ab} R_{cd} R^{cd}
\nn\\
& + 8 (d - 2)^2 (d -1) (8160 - 19934 d + 18411 d^2 - 6271 d^3 - 1872 d^4 + 2790 d^5 
\nn\\
&- 1261 d^6 + 301 d^7 - 38 d^8 + 2 d^9) R_{a}{}^{c} R^{ab} R_{bc} R 
+ 2 (374400 - 1072928 d 
\nn\\
&+ 1257694 d^2 - 724744 d^3 + 156052 d^4 + 53793 d^5 - 49657 d^6 + 17344 d^7 
\nn\\
&- 3698 d^8 + 525 d^9 - 47 d^{10} + 2 d^{11}) R_{ab} R^{ab} R^2 + 24 (d - 2) (-128640 + 368958 d 
\nn\\
&- 429005 d^2 + 239408 d^3 - 43691 d^4 - 22101 d^5 + 15982 d^6 - 4406 d^7 + 625 d^8 
\nn\\
&- 43 d^9 + d^{10}) R^{ab} R^{cd} R R_{acbd} - 3 (361600 - 1116656 d + 1410902 d^2 
\nn\\
&- 875630 d^3 + 208502 d^4 + 51581 d^5 - 38382 d^6 - 577 d^7 + 6668 d^8 - 2637 d^9 
\nn\\
&+ 500 d^{10} - 49 d^{11} + 2 d^{12}) R^2 R_{abcd} R^{abcd} - 24 (d - 2) (d -1) (-119680 
\nn\\
&+ 338440 d - 401078 d^2 + 240034 d^3 - 58237 d^4 - 13906 d^5 + 14831 d^6 
\nn\\
&- 4890 d^7 + 849 d^8 - 78 d^9 + 3 d^{10}) R_{a}{}^{c} R^{ab} R^{de} R_{bdce} \nn\\
&+ 6 (d - 2)^2 (d -1) (28160 - 110076 d + 172418 d^2 - 146251 d^3 
\nn\\
&+ 75674 d^4 - 25778 d^5 + 6287 d^6 - 1211 d^7 + 188 d^8 - 20 d^9 + d^{10}) R^{ab} R^{cd} R_{ac}{}^{ef} R_{bdef} 
\nn\\
&+ 2 (d - 3) (d - 2)^2 (d -1) (-2400 + 9201 d - 9929 d^2 + 2690 d^3 + 1954 d^4 
\nn\\
&- 1667 d^5 + 507 d^6 - 72 d^7 + 4 d^8) R R_{ab}{}^{ef} R^{abcd} R_{cdef} + 3 (-113920 
\nn\\
&+ 801792 d - 1837992 d^2 + 2067094 d^3 - 1242116 d^4 + 346968 d^5 + 4985 d^6 \nn\\
&- 18628 d^7 - 8905 d^8 + 9138 d^9 - 3089 d^{10} + 546 d^{11} - 51 d^{12} + 2 d^{13}) R_{ab} R^{ab} R_{cdef} R^{cdef} 
\nn\\
&- 6 (d - 3) (d - 2)^2 (d -1) (-22 + 26 d - 9 d^2 + d^3) (320 - 709 d + 588 d^2 
\nn\\
&- 292 d^3 + 106 d^4 - 23 d^5 + 2 d^6) R^{ab} R_{a}{}^{cde} R_{bc}{}^{fh} R_{defh} \big] 
\nn\\
&+ R_{a}{}^{e}{}_{c}{}^{f} R^{abcd} R_{b}{}^{h}{}_{e}{}^{j} R_{dhfj}
\end{align}

\begin{align}
\mathcal{S}_d^{(2)} &= \frac{1}{3(d - 3)^2 (d - 2)^2 (d -1) d (11 - 6 d + d^2) (19 - 18 d + 3 d^2) (-22 + 26 d - 9 d^2 + d^3)} \times
\nn\\
&\times\big[
-2 (d - 2)^2 (-578688 + 2025158 d - 3185710 d^2 + 2977426 d^3 - 1839784 d^4
\nn\\
& + 791721 d^5 - 244086 d^6 + 54763 d^7 - 8972 d^8 + 1049 d^9 - 80 d^{10} + 3 d^{11}) R_{a}{}^{c} R^{ab} R_{b}{}^{d} R_{cd} 
\nn\\
&+ (2281872 - 8031408 d + 12067376 d^2 - 9693872 d^3 + 3903996 d^4 + 22113 d^5 \nn\\
&- 946024 d^6 + 572163 d^7 - 189362 d^8 + 39599 d^9 - 5244 d^{10} + 405 d^{11} - 14 d^{12}) R_{ab} R^{ab} R_{cd} R^{cd} 
\nn\\
&+ 8 (d - 2)^2 (d -1) (-6987 + 21346 d - 29861 d^2 + 26093 d^3 - 15931 d^4 + 6882 d^5 
\nn\\
&- 2031 d^6 + 385 d^7 - 42 d^8 + 2 d^9) R_{a}{}^{c} R^{ab} R_{bc} R + 2 (-320580 + 1132666 d 
\nn\\
&- 1781245 d^2 + 1646682 d^3 - 998922 d^4 + 421855 d^5 - 128958 d^6 + 29348 d^7 
\nn\\
&- 5024 d^8 + 627 d^9 - 51 d^{10} + 2 d^{11}) R_{ab} R^{ab} R^2 - 12 (d - 2) (-220296 
\nn\\
&+ 774954 d - 1199885 d^2 + 1070366 d^3 - 604828 d^4 + 223750 d^5 - 53844 d^6 \nn\\
&+ 7998 d^7 - 628 d^8 + 12 d^9 + d^{10}) R^{ab} R^{cd} R R_{acbd} - 3 (-309620 + 1149158 d 
\nn\\
&- 1856955 d^2 + 1675917 d^3 - 875073 d^4 + 209908 d^5 + 40520 d^6 - 53295 d^7 \nn\\
&+ 21313 d^8 - 4912 d^9 + 693 d^{10} - 56 d^{11} + 2 d^{12}) R^2 R_{abcd} R^{abcd} 
\nn\\
&- 24 (d - 2) (d -1) (102476 - 371148 d + 606224 d^2 - 585295 d^3 + 368632 d^4 
\nn\\
&- 157824 d^5 + 46423 d^6 - 9251 d^7 + 1194 d^8 - 90 d^9 + 3 d^{10}) R_{a}{}^{c} R^{ab} R^{de} R_{bdce} 
\nn\\
&+ 6 (d - 2)^2 (d -1) (-24112 + 104237 d - 184591 d^2 + 177665 d^3 - 102275 d^4 
\nn\\
&+ 35933 d^5 - 7258 d^6 + 601 d^7 + 55 d^8 - 16 d^9 + d^{10}) R^{ab} R^{cd} R_{ac}{}^{ef} R_{bdef} 
\nn\\
&+ (d - 3) (d - 2)^2 (d -1) (4110 - 23613 d + 44912 d^2 - 42687 d^3 + 23334 d^4 
\nn\\
&- 7715 d^5 + 1532 d^6 - 169 d^7 + 8 d^8) R R_{ab}{}^{ef} R^{abcd} R_{cdef} + 3 (97544 - 765604 d 
\nn\\
&+ 2080704 d^2 - 2942717 d^3 + 2459345 d^4 - 1222083 d^5 + 288468 d^6 + 44796 d^7 
\nn\\
&- 62477 d^8 + 24383 d^9 - 5446 d^{10} + 743 d^{11} - 58 d^{12} + 2 d^{13}) R_{ab} R^{ab} R_{cdef} R^{cdef} 
\nn\\
&- 6 (d - 3) (d - 2)^2 (d -1) (-22 + 26 d - 9 d^2 + d^3) (-274 + 409 d - 67 d^2 
\nn\\
&- 161 d^3 + 103 d^4 - 24 d^5 + 2 d^6) R^{ab} R_{a}{}^{cde} R_{bc}{}^{fh} R_{defh} \big]
\nn\\
&+  R_{a}{}^{e}{}_{c}{}^{f} R^{abcd} R_{b}{}^{h}{}_{d}{}^{j} R_{ehfj}
\end{align}

\begin{align}
\mathcal{S}_d^{(3)} &= \frac{1}{3 (d - 3)^2 (d - 2) (d -1) d (11 - 6 d + d^2) (19 - 18 d + 3 d^2) (-22 + 26 d - 9 d^2 + d^3)} \times
\nn\\
&\times \big[ 
-4 (d - 2) (-718080 + 2405582 d - 3666144 d^2 + 3359133 d^3 - 2057938 d^4 
\nn\\
&+ 887142 d^5 - 276120 d^6 + 62662 d^7 - 10296 d^8 + 1182 d^9 - 86 d^{10} + 3 d^{11}) R_{a}{}^{c} R^{ab} R_{b}{}^{d} R_{cd} 
\nn\\
&- 4 (707880 - 2115012 d + 2700668 d^2 - 1809780 d^3 + 561468 d^4 + 61133 d^5 - 134394 d^6 
\nn\\
&+ 60426 d^7 - 15005 d^8 + 2238 d^9 - 189 d^{10} + 7 d^{11}) R_{ab} R^{ab} R_{cd} R^{cd} 
\nn\\
&+ 16 (d - 2) (d -1) (-8670 + 30262 d - 47247 d^2 + 43299 d^3 - 25747 d^4 + 10271 d^5 
\nn\\
&- 2734 d^6 + 466 d^7 - 46 d^8 + 2 d^9) R_{a}{}^{c} R^{ab} R_{bc} R + 4 (198900 - 592178 d + 790224 d^2 
\nn\\
&- 617415 d^3 + 313537 d^4 - 109500 d^5 + 27237 d^6 - 4900 d^7 + 624 d^8 - 51 d^9 + 2 d^{10}) R_{ab} R^{ab} R^2 
\nn\\
&+ 48 (d - 2) (-68340 + 203532 d - 268574 d^2 + 203038 d^3 - 95967 d^4 + 29190 d^5 
\nn\\
&- 5665 d^6 + 667 d^7 - 42 d^8 + d^9) R^{ab} R^{cd} R R_{acbd} - 6 (192100 - 603774 d + 820554 d^2 
\nn\\
&- 605255 d^3 + 237492 d^4 - 22951 d^5 - 24843 d^6 + 14329 d^7 - 3890 d^8 + 609 d^9 - 53 d^{10} 
\nn\\
&+ 2 d^{11}) R^2 R_{abcd} R^{abcd} - 48 (d - 2) (d -1) (-63580 + 183572 d - 244118 d^2 
\nn\\
&+ 192444 d^3 - 97734 d^4 + 32893 d^5 - 7308 d^6 + 1032 d^7 - 84 d^8 + 3 d^9) R_{a}{}^{c} R^{ab} R^{de} R_{bdce}
\nn\\
& + 12 (d - 2) (d -1) (-29920 + 120000 d - 196892 d^2 + 175930 d^3 - 93864 d^4 
\nn\\
&+ 30115 d^5 - 5212 d^6 + 193 d^7 + 99 d^8 - 18 d^9 + d^{10}) R^{ab} R^{cd} R_{ac}{}^{ef} R_{bdef} 
\nn\\
&+ 4 (d - 3) (d - 2) (d -1) (2550 - 15414 d + 28633 d^2 - 26167 d^3 + 13715 d^4 
\nn\\
&- 4351 d^5 + 830 d^6 - 88 d^7 + 4 d^8) R R_{ab}{}^{ef} R^{abcd} R_{cdef} + 6 (-60520 + 414664 d 
\nn\\
&- 945458 d^2 + 1097752 d^3 - 719367 d^4 + 242784 d^5 - 3125 d^6 - 36155 d^7 + 17569 d^8 
\nn\\
&- 4430 d^9 + 659 d^{10} - 55 d^{11} + 2 d^{12}) R_{ab} R^{ab} R_{cdef} R^{cdef}
\nn\\
& - 12 (d - 3) (d - 2) (d -1) (-22 + 26 d - 9 d^2 + d^3) (-340 + 494 d - 70 d^2 - 185 d^3 
\nn\\
&+ 112 d^4 - 25 d^5 + 2 d^6) R^{ab} R_{a}{}^{cde} R_{bc}{}^{fh} R_{defh} \big]
\nn\\
&+  R_{ab}{}^{ef} R^{abcd} R_{ce}{}^{hj} R_{dfhj}
\end{align}

\begin{align}
\mathcal{S}_d^{(4)} &= \frac{1}{3 (d - 3)^2 (d - 2) (d -1) d (11 - 6 d + d^2) (19 - 18 d + 3 d^2) (-22 + 26 d - 9 d^2 + d^3)} \times
\nn\\
&\times \big[
-8 (d - 2) (-718080 + 2405582 d - 3666144 d^2 + 3359133 d^3 - 2057938 d^4
\nn\\
& + 887142 d^5 - 276120 d^6 + 62662 d^7 - 10296 d^8 + 1182 d^9 - 86 d^{10} + 3 d^{11}) R_{a}{}^{c} R^{ab} R_{b}{}^{d} R_{cd} 
\nn\\
&- 8 (707880 - 2115012 d + 2700668 d^2 - 1809780 d^3 + 561468 d^4 + 61133 d^5 \nn\\
&- 134394 d^6 + 60426 d^7 - 15005 d^8 + 2238 d^9 - 189 d^{10} + 7 d^{11}) R_{ab} R^{ab} R_{cd} R^{cd}
\nn\\
& + 32 (d - 2) (d -1) (-8670 + 30262 d - 47247 d^2 + 43299 d^3 - 25747 d^4 + 10271 d^5 
\nn\\
&- 2734 d^6 + 466 d^7 - 46 d^8 + 2 d^9) R_{a}{}^{c} R^{ab} R_{bc} R + 8 (198900 - 592178 d + 790224 d^2 
\nn\\
&- 617415 d^3 + 313537 d^4 - 109500 d^5 + 27237 d^6 - 4900 d^7 + 624 d^8 - 51 d^9 + 2 d^{10}) R_{ab} R^{ab} R^2 
\nn\\
&+ 96 (d - 2) (-68340 + 203532 d - 268574 d^2 + 203038 d^3 - 95967 d^4 + 29190 d^5 
\nn\\
&- 5665 d^6 + 667 d^7 - 42 d^8 + d^9) R^{ab} R^{cd} R R_{acbd} - 12 (192100 - 603774 d + 820554 d^2 
\nn\\
&- 605255 d^3 + 237492 d^4 - 22951 d^5 - 24843 d^6 + 14329 d^7 - 3890 d^8 + 609 d^9 - 53 d^{10} 
\nn\\
&+ 2 d^{11}) R^2 R_{abcd} R^{abcd} - 96 (d - 2) (d -1) (-63580 + 183572 d - 244118 d^2 
\nn\\
&+ 192444 d^3 - 97734 d^4 + 32893 d^5 - 7308 d^6 + 1032 d^7 - 84 d^8 + 3 d^9) R_{a}{}^{c} R^{ab} R^{de} R_{bdce} 
\nn\\
&+ 24 (d - 2) (d -1) (-29920 + 120000 d - 196892 d^2 + 175930 d^3 - 93864 d^4 
\nn\\
&+ 30115 d^5 - 5212 d^6 + 193 d^7 + 99 d^8 - 18 d^9 + d^{10}) R^{ab} R^{cd} R_{ac}{}^{ef} R_{bdef} 
\nn\\
&+ 8 (d - 3) (d - 2) (d -1) (2550 - 15414 d + 28633 d^2 - 26167 d^3 + 13715 d^4 
\nn\\
&- 4351 d^5 + 830 d^6 - 88 d^7 + 4 d^8) R R_{ab}{}^{ef} R^{abcd} R_{cdef} + 12 (-60520 + 414664 d 
\nn\\
&- 945458 d^2 + 1097752 d^3 - 719367 d^4 + 242784 d^5 - 3125 d^6 - 36155 d^7 + 17569 d^8 
\nn\\
&- 4430 d^9 + 659 d^{10} - 55 d^{11} + 2 d^{12}) R_{ab} R^{ab} R_{cdef} R^{cdef} 
\nn\\
&- 24 (d - 3) (d - 2) (d -1) (-22 + 26 d - 9 d^2 + d^3) (-340 + 494 d - 70 d^2 
\nn\\
&- 185 d^3 + 112 d^4 - 25 d^5 + 2 d^6) R^{ab} R_{a}{}^{cde} R_{bc}{}^{fh} R_{defh} \big]
\nn\\
&+ R_{ab}{}^{ef} R^{abcd} R_{cd}{}^{hj} R_{efhj}
\end{align}

The following two Lagrangian densities are relevant only for the four dimensional theory.

\begin{align}
\mathcal{S}_4^{(5)} &= - \frac{14}{5} R_{ab} R^{ab} R_{cd} R^{cd} -  \frac{20}{3} R_{a}{}^{b} R_{b}{}^{c} R_{c}{}^{d} R_{d}{}^{a} -  \frac{8}{5} R^{ac} R^{bd} R R_{abcd} 
\nn\\
&+ \frac{104}{5} R^{ab} R_{e}{}^{d} R^{ec} R_{acbd} + R_{ef} R^{ef} R_{abcd} R^{abcd} + \frac{1}{5} R^2 R_{abcd} R^{abcd} 
\nn\\
&-  \frac{56}{15} R^{ab} R_{cd}{}^{h}{}_{a} R^{cdef} R_{efhb} + R_{abc}{}^{e} R^{abcd} R_{fhjd} R^{fhj}{}_{e}
\end{align}

\begin{align}
\mathcal{S}_4^{(6)} &= - \frac{308}{15} R_{ab} R^{ab} R_{cd} R^{cd} -  \frac{64}{3} R_{a}{}^{b} R_{b}{}^{c} R_{c}{}^{d} R_{d}{}^{a} + \frac{64}{15} R^{ac} R^{bd} R R_{abcd} + \frac{1088}{15} R^{ab} R_{e}{}^{d} R^{ec} R_{acbd}
\nn\\
& + \frac{28}{3} R_{ef} R^{ef} R_{abcd} R^{abcd} -  \frac{8}{15} R^2 R_{abcd} R^{abcd} -  \frac{224}{15} R^{ab} R_{cd}{}^{h}{}_{a} R^{cdef} R_{efhb} 
\nn\\
&+ R_{abcd} R^{abcd} R_{fhje} R^{fhje}
\end{align}


\bibliography{LBIB}
\bibliographystyle{JHEP}
\end{document}